\documentclass{aa} % for the letters 
\usepackage{graphics,graphicx}
\usepackage[varg]{txfonts}
\usepackage{rotating}
\usepackage{lscape}
\usepackage{natbib}
\newcommand{\asec}{$^{\prime\prime}$}

\def\i{I05345}

\def\H{N$_{2}$H$^{+}$}
\def\D{N$_{2}$D$^{+}$}

\def\AMM{NH$_3$}
\def\AMMtoH{NH$_3$/N$_{2}$H$^{+}$}

\def\CO{\mbox{$^{12}$CO}}

\def\HII{H{\sc ii}}

\def\kms{\mbox{km~s$^{-1}$}}
\def\cmc{cm$^{-3}$}
\def\cmq{cm$^{-2}$}

\def\solm{\mbox{M$_\odot$}}

\def\Tex{\mbox{$T_{\rm ex}$}}

\def\Tr{\mbox{$T_{\rm rot}$}}

\def\deltav{\mbox{$\Delta V$}}

\begin{document}

\title{Temperature and kinematics of protoclusters with intermediate and high-mass stars: the case of IRAS 05345+3157
}
%\subtitle{...}
\author{F. Fontani \inst{1} 
             %\and G. Busquet \inst{3} 
             \and  P. Caselli \inst{2}
           \and Q. Zhang \inst{3}   
        \and J. Brand \inst{4}  
         \and G. Busquet \inst{5} 
         \and Aina Palau \inst{6} 
        }
\offprints{F. Fontani, \email{fontani@arcetri.astro.it}}
\institute{INAF - Osservatorio Astrofisico di Arcetri, L.go E. Fermi 5, I-50125, Firenze, Italy
%	   \and
%             Institut de Radio-Astronomie Millim\'etrique, 300 rue de la Piscine, Domaine Universitaire, 38406 Saint Martin  d'H\`eres, France
	\and
	   School of Physics and Astrophysics, University of Leeds, Leeds, LS2 9JT, UK
	    \and
	    Harvard-Smithsonian Center for Astrophysics, 60 Garden Street MS78, Cambridge, MA 02138, USA
	   \and
	   INAF - Istituto di Radioastronomia, via P. Gobetti 101, I-40129 Bologna, Italy
	  \and
	  INAF-Istituto di Astrofisica e Planetologia Spaziali, via Fosso del Cavaliare 100, 00133 Roma, Italy
	  \and
	  Institut de Ci\`encies de l'Espai (CSIC-IEEC), Campus UAB-Facultat de Ci\`encies, Torre C5-parell 2, 08193, Bellaterra, Spain
	   } 
\date{Received date; accepted date}

%\markboth{Fontani et al.: Mol160}{}
\titlerunning{The protocluster in IRAS 05345+3157}
\authorrunning{Fontani et al.}

\abstract
{Improving our understanding of the complex star formation process in clusters requires studies of 
star-forming clouds to search for dependencies of the physical properties on environmental variables, 
such as overall density, stellar crowding and feedback from massive (proto-)stars.}
{We aim to map at small spatial scales the temperature and the velocity field in 
the protocluster associated with IRAS 05345+3157, which contains both 
intermediate-/high-mass protostellar candidates and starless condensations, and is thus an excellent
location to investigate the role of massive protostars on protocluster evolution.}
{We observed the ammonia (1,1) and (2,2) inversion transitions with the VLA.
Ammonia is the best thermometer for dense and cold gas, and the observed transitions
have critical densities able to trace the kinematics of the intracluster gaseous medium.}
{The ammonia emission is extended and distributed in two filamentary structures.
The starless condensations are colder than the star-forming cores,
but the gas temperature across the whole protocluster is higher (by a factor of ~1.3--1.5)
than that measured typically in both infrared dark clouds and low-mass
protoclusters. The non-thermal contribution to the observed line broadening is at least
a factor of 2 larger than the expected thermal broadening even in starless condensations,
contrary to the close-to-thermal line widths measured in low-mass quiescent dense cores. 
The \AMMtoH\ abundance ratio is greatly
enhanced (a factor of 10) in the pre--stellar core candidates, probably
due to freeze-out of most molecular species heavier than He. }
{The more massive and evolved objects likely play a dominant
role in the physical properties and kinematics of the protocluster.
The high level of turbulence and the fact that the measured core masses
are larger than the expected thermal Jeans masses indicate that
turbulence likely was an important factor in the initial fragmentation of the parental
clump and can provide support against further fragmentation of the cores. 
}
%Higher resolution studies of the large-scale gas
%kinematics and comparison with hydrodynamic models would help to highlight this.}

\maketitle
%
%________________________________________________________________

\section{Introduction}
\label{Introduction}

Most stars form in clusters in the densest portions of giant molecular clouds, 
but the initial conditions of the star formation process in clusters are still poorly
understood. Studies have unveiled the physical and chemical 
properties of isolated low-mass pre--stellar cores, i.e. starless condensations on 
the verge of gravitational collapse, in nearby low-mass star-forming regions.
These works show that pre--stellar cores
have dense ($n\sim 10^5 - 10^6$ \cmc ) and cold ($T\sim 10$ K) nuclei, and that 
their internal motions are thermally dominated, as demonstrated
by their close-to-thermal line widths, even when observed at low angular 
resolution (see Bergin \& Tafalla~\citeyear{bergin} for a review).
%In the cold and dense nuclei of these cores,  C-bearing molecular species such 
%as CO and CS are strongly depleted (e.g.~Caselli et al.~\citeyear{caselli02})
%%Tafalla et al.~\citeyear{tafalla02}),
%while N-bearing species such as \H\ and \AMM\ maintain
%large abundances in the gas phase (Caselli et al.~\citeyear{caselli02}; Crapsi et al.~\citeyear{crapsi}).
One as yet poorly investigated major issue is if, and how, the environment 
influences the physical and chemical properties
of these pre--stellar cores. In clusters containing several forming 
stars in a small area ($\leq 0.1$ pc), turbulence, relative motions, and 
interactions with nearby forming (proto-)stars can in principle affect the
less evolved condensations (Ward-Thompson et al.~\citeyear{wardthompson}). 
Such interaction is expected to be very important in
clusters containing intermediate- and high-mass protostars or newly
formed massive stars, given the typical energetic feedback provided
by these objects (powerful outflows, strong winds, UV radiation, expanding \HII\ regions).
There is also vigorous theoretical debate on how star formation proceeds in  
clustered regions. Specifically: is the local kinematics of the gas dominated by  
feedback from protostellar outflows of already-forming, generally low-mass stars 
(Nakamura \& Li~\citeyear{nakamura}) or by feedback from high-mass stars?  
To what extent does turbulence regulate the star formation rate? 
Can a  model of star formation starting from quiescent starless cores be applied to clustered regions 
(McKee \& Tan~\citeyear{mcKee})? To put constraints on theoretical models it is
crucial to determine the kinematics and the physical properties 
of starless cores in clustered environments, especially those containing sources 
of very energetic feedback.

An observational effort in this direction
has started, but it is mostly concentrated on nearby low-mass star-forming 
regions like Ophiuchus (see e.g.~Andr\'e et al.~\citeyear{andre}; 
Friesen et al.~\citeyear{friesen}), Perseus (Foster et al.~\citeyear{foster}) and the Pipe
Nebula (Rathborne et al.~\citeyear{rathborne}, Forbrich et al.~\citeyear{fobrich}). 
Foster et al.~(\citeyear{foster}) and Friesen et al.~(\citeyear{friesen}) show that 
cores within low-mass star-forming clusters have typically 
higher kinetic temperatures ($\sim 15$~K) than low-mass isolated ones (10~K). 
On the other hand, the kinematics seems to be dominated by thermal 
motions like in more isolated cores,  even though the external environment is 
turbulent (Andr\'e et al.~\citeyear{andre}, Rathborne et al~\citeyear{rathborne}). 
Few works have been performed towards protoclusters
forming intermediate- and high-mass stars to date (Palau et al.~\citeyear{palau2007a},~\citeyear{palau07b},
Leurini et al.~\citeyear{leurini},
Pillai et al.~\citeyear{pillai11}), due to the fact that the typically large 
distances ($\geq 0.5 - 1$~kpc) of high-mass star-forming regions make the 
study of clustered environments challenging. 
%Moreover, these works are
%mainly focussed on the properties of the protostellar objects rather
%then on those of the starless condensations possibly present
%(e.g.~Palau et al.~\citeyear{palau2007a},~\citeyear{palau07b}, Leurini et al.~\citeyear{leurini}). 
%The few studies of (proto-)clusters containing intermediate- or high-mass forming 
%stars provide contrasting results: both high and low levels of turbulence
%are measured (paperII, 
%Lee et al.~\citeyear{lee} for IRAS 05345+3157; Wang et al.~\citeyear{wang};
%for G28.34+0.06; Beuther et al.~(\citeyear{beuther09}), and 
%kinetic temperatures are comparable to the values measured in 
%clustered cores in Perseus and Ophiuchus (e.g. Palau et al.~\citeyear{palau07b} 
%for IRAS 20293+3952). 
%Beuther et al.~(\citeyear{beuther09}) found relatively 
%low levels of turbulence in the starless cores close to the high-mass star-forming
%region IRAS 19175+1357, but given the linear projected separation between
%the starless cores and the centre of star formation activity (approximately 0.3~pc),
%these might not interact.
%However, 
%to understand whether the results obtained in the few examples 
%shown above can be considered 'typical', 
%To shed light on the clustered star formation process in the whole Galaxy,
%one must increase the observations of protoclusters including both starless 
%and massive star-forming condensations, and investigate at high angular
%resolution the properties of the dense gas (the cradle of protostars) associated 
%with them.

The target of the present paper is a well-studied protocluster embedded
in a massive (180 $M_{\odot}$) pc-scale clump 
(Fontani et al.~\citeyear{fontani06}, Klein et al.~\citeyear{klein}). 
The protocluster harbours intermediate- to high-mass
protostellar objects and starless condensations 
(Fontani et al.~\citeyear{fontani08}, Fontani et al.~\citeyear{fontani09}, hereafter 
paper~I and paper~II, respectively), 
and is located at a distance of 1.8~kpc, $\sim 1$\arcmin\ north-east from the
IRAS source 05345+3157 associated with
a rich cluster of infrared sources (Varricatt et al.~\citeyear{varricatt}). Hereafter, the 
protocluster will be called I05345. One can summarise 
the main findings of the previous observations as follows
(as general reference, see Fig.~\ref{I05345_NH311_int}):
\begin{itemize}
\item at a resolution of $\sim 3$\arcsec , the millimetre continuum
reveals three dusty cores, C1, C2 and C3. 
C1 is then resolved into two peaks, C1-a and C1-b, down to
$1\farcs 5$ resolution (paper~I, paper~II);
\item  both C1-a and C1-b are associated with infrared emission and are
believed to harbour intermediate- to high-mass protostars, with C1-b
being the most massive and associated with a hot core, unlike
C1-a; in paper~II we wrongly claimed that C1-b is likely associated with 
3.6~cm-continuum emission (Molinari et al.~\citeyear{mol02}) due to an
error in the coordinates of the peak. The radio continuum emission is instead more 
likely associated with C2, and it can be due either to a radio jet or to
a deeply embedded \HII\ region. In either
case, C2 harbours a very young intermediate- or high-mass object; 
C3 is also associated with both dense gas and an infrared source, but its nature
is less clear (paper~I, Varricatt et al.~\citeyear{varricatt}, Lee et al.~\citeyear{lee}); 
\item an extended gaseous emission detected in \H\ (1--0) with the PdBI 
encompasses all continuum cores and reveal additional condensations
(paper~I, paper~II);
\item two of these condensations, called N and S, were also detected in emission 
lines of \D\ but not in the mm- and IR-continuum. Based on the chemistry
and on the lack of infrared emission, N and S are probably
starless and in the pre--stellar phase (paper~I);
\item a widespread CO outflow likely driven by C1-b and/or C1-a
interacts with the starless condensations (paper~II). 
\end{itemize}
These features make this protocluster an excellent location to investigate if and how 
starless cores are influenced by the feedback from forming intermediate- to high-mass stars.
In this paper we present new VLA high-angular resolution observations of the (1,1)
and (2,2) inversion transitions of \AMM\ towards I05345, focusing mainly 
on the kinetic temperature and the kinematics of the dense gas. 
\AMM\ is the ideal thermometer because
it is known to be a tracer of relatively dense gas and to be present in the gas phase
even in regions where CO and other C-bearing species are frozen
onto dust grains, and the \AMM\ (2,2) to (1,1) line
ratio is sensitive to temperature. Also, the critical
density of both transitions (about $10^{4}$ \cmc ) allows us to investigate
the kinematics of the gaseous envelope in which the dense protocluster
cores are embedded. 

In Sect. \ref{obs} we outline the
details of the VLA observations and data reduction. Sect.~\ref{res} presents the immediate 
observational results and the method used 
to derive the physical parameters from \AMM , while in Sect.~\ref{sect_maps} we
show the maps of these parameters. In Sect.~\ref{discu} we discuss 
some aspects of the chemistry and the implications of this work in the context of the
protocluster formation problem. A summary and the main conclusions are given in 
Sect.~\ref{summary}.

\section{Observations and data reduction}
\label{obs}

The NRAO Very Large Array (VLA)\footnote{The VLA is operated by the
National Radio Astronomy Observatory (NRAO), a facility of the
National Science Foundation operated under cooperative agreement
by Associated Universities, Inc.} ammonia observations towards I05345 were performed on 
October 31 and November 2, 2009. As phase centre we used the nominal position of the sub-millimetre 
peak detected with the JCMT (Fontani et al.~2006), namely 
RA(J2000) = 05$^{\rm h}$37$^{\rm m}$$52\rm \fs 4$ and 
Dec(J2000) = 32$^{\circ}$00$^{\prime}$06$^{\prime \prime }$, which 
roughly corresponds to the position of C1. 
The local standard of rest velocity $V_{\rm LSR}$ is --18.4 \kms . 
The \AMM\ ($J$,$K$) = (1,1) and (2,2) inversion transitions at 23.694496 and 23.722633 GHz, 
respectively, were observed simultaneously, 
using the 2-IF spectral line mode of the correlator, with 3.125 MHz bandwidths: the first IF 
covered the main line and the two innermost satellites of the (1,1) transition, while the other
IF was used to observe the (2,2) line. Channel spacing was 24.414 kHz, correspoding
to a spectral resolution of $\sim 0.3$ \kms\ at the frequency of the lines. 
The array was used in the most compact configuration (D), offering baselines from 35 m to 1 km. 
The primary beam was $\sim 2$\arcmin\ at the line frequencies.
The flux density scale was established by observing the standard primary calibrators
3C286 and 3C48, and the uncertainty is expected to be 
less than $\sim 15\%$. Gain calibration was ensured by frequent observations of the compact
quasar J0555+398.
%with a measured flux density at the time of the observations of 3.4 Jy. 
The quasar 3C84 was used for passband calibration.
Pointing corrections were derived from X-band observations of nearby quasars and applied online.
The data were edited and calibrated following standard tasks and procedures
of the Astronomical Image Processing System (AIPS). Imaging and deconvolution were 
performed using the IMAGR task, applying natural weighting to the visibilities. 
The resulting synthesised beam full widths at half-maximum are 
$2\farcs 33 \times 2\farcs 27$ (position angle $92^{\circ}$) and 
$2\farcs 20 \times 2\farcs 16$ (position angle $90^{\circ}$) for the (1,1) 
and (2,2) lines, respectively. 
The noise level in each channel is about 1.4 mJy beam$^{-1}$ and 
1.5 mJy beam$^{-1}$ for the (1,1) and (2,2) data, respectively. 

Calibrated channel maps and spectra were analysed using the 
GILDAS\footnote{The GILDAS software is available at http://www.iram.fr/ IRAMFR/GILDAS} 
software developed at IRAM and the Observatoire de Grenoble. 

\section{Immediate results and derivation of physical parameters}
\label{res}

\begin{figure*}
\centerline{\includegraphics[angle=-90,width=16cm]{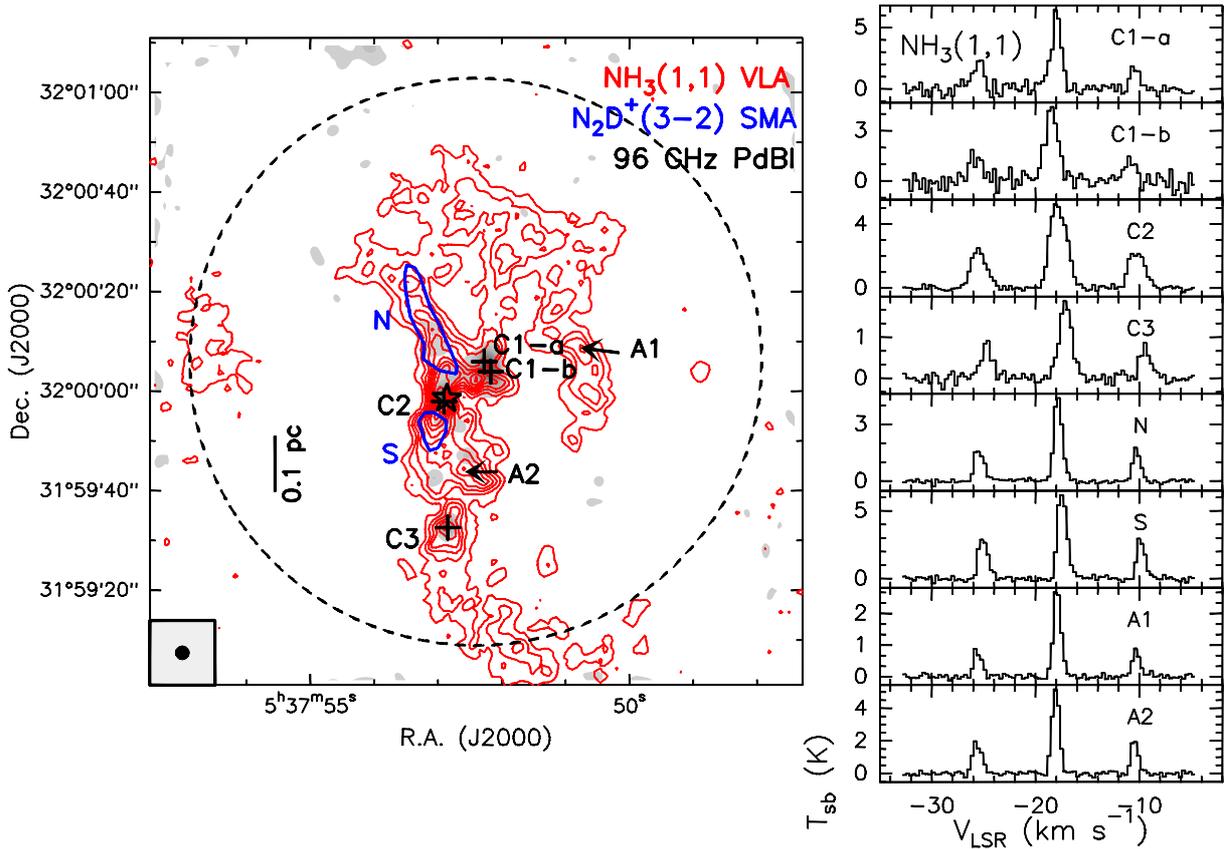}}
\caption{Left panel: zero-th order moment map of the \AMM\ (1,1) line 
obtained with the VLA towards \i\ (red contours). Contours start 
from the 6$\%$ of the maximum (0.039 Jy beam$^{-1}$ \kms ), and are in steps of 10$\%$ 
of the maximum (0.065 Jy beam$^{-1}$ \kms ). The integration velocity interval is
--19.33 and --16.24 \kms , corresponding
to the main group of hyperfine components. Contour levels start from the 
3$\sigma$ rms ($\sim 0.0012$ Jy beam$^{-1}$ \kms ), and are in steps of 3 $\sigma$.
The grey scale represents the 96 GHz continuum observed with the PdBI 
(first level is the 3$\sigma$ rms = 0.004 Jy beam$^{-1}$; 
step = 3$\sigma$ rms), 
and the compact 3~millimetre continuum cores detected in paper~II
are marked by crosses and labelled as C1-a, C1-b, C2 and C3. 
The two blue contours correspond to the 3$\sigma$ level of the 
integrated emission of \D\ (3--2) observed with the SMA
(paper~I). The arrows identify the
emission peaks of the ammonia cores A1 and A2 (this work). The star
pinpoints the 3.6~cm continuum peak detected by Molinari et al.~(\citeyear{mol02}).
The dashed circle represents the 
VLA primary beam at $\sim 22$ GHz ($\sim 114$\asec ). 
The ellipse in the bottom left corner shows the synthesised beam of 
the \AMM\ (1,1) image. 
\newline
Right panel: Spectra of the \AMM\ (1,1) line towards the continuum cores
C1-a, C1-b, C2 and C3, as well as the \D\ condensations N and S, and
the ammonia cores A1 and A2
(see left panel) in synthesised beam temperature units ($T_{\rm sb}$). 
The spectra of C1-a and C1-b have been averaged over the 3$\sigma$~rms
contour level of the 284~GHz emission shown in Fig.~3 of paper~II, because
these are resolved in that image only.
For C2 and C3, we used the 225~GHz 3$\sigma$~rms
contour level shown in Fig.~1 of paper~I, because the 225~GHz
image is less affected by the flux filtering problem.}
\label{I05345_NH311_int}
\end{figure*}

\begin{figure*}
\centerline{\includegraphics[angle=-90,width=9cm]{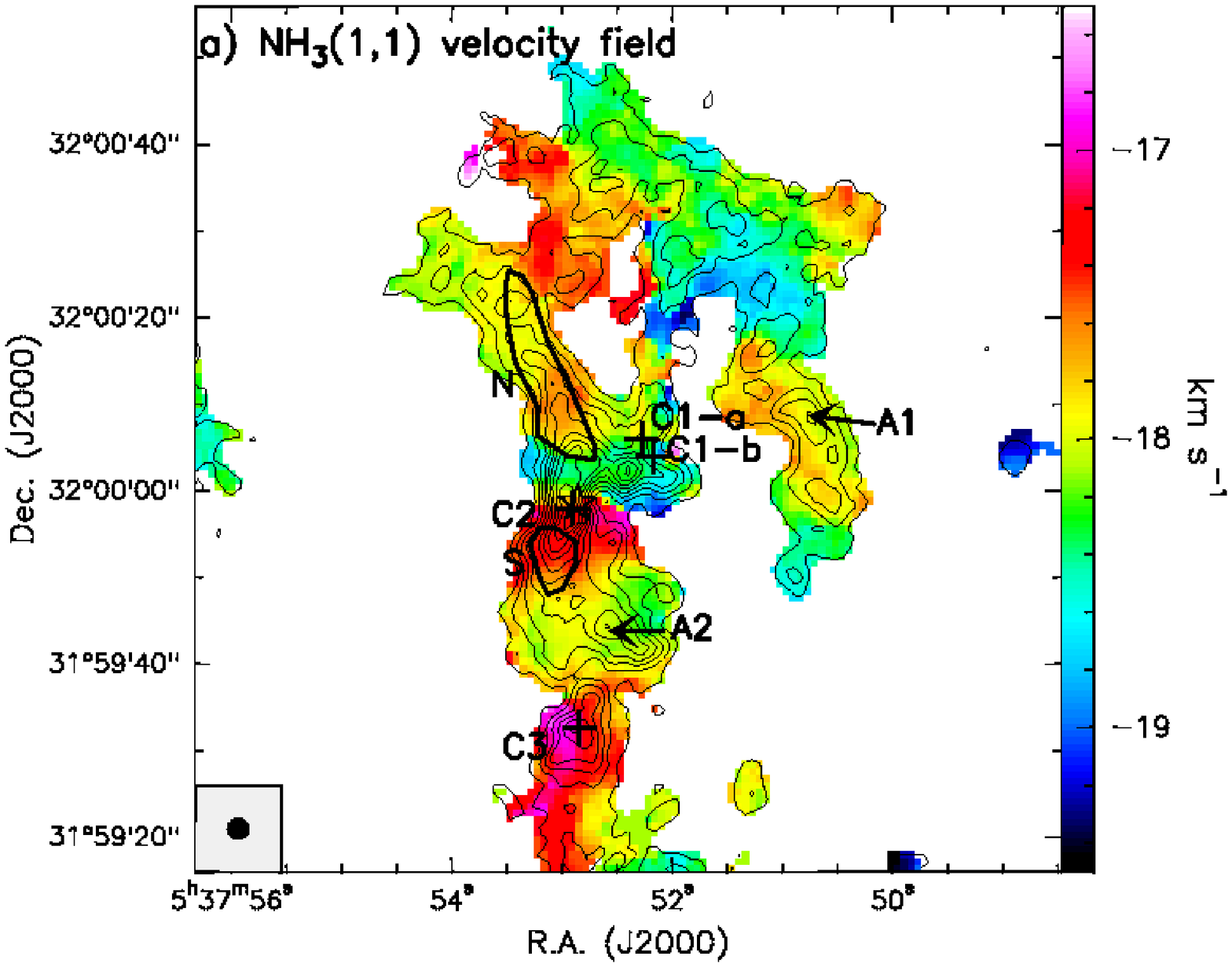}
                    \includegraphics[angle=-90,width=9cm]{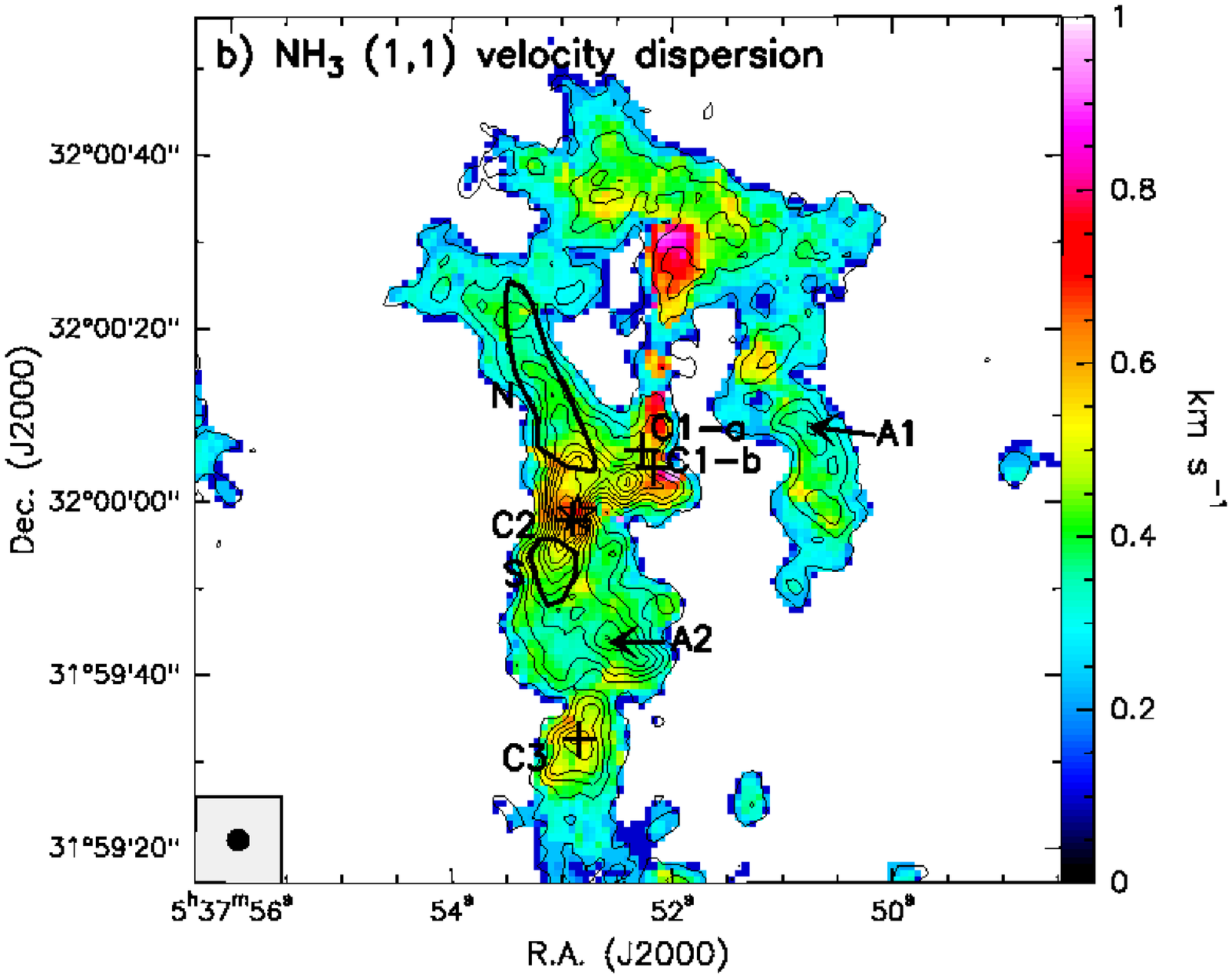}}
\caption{(a) first-order moment map (colour scale) of the \AMM\ (1,1) main group
of hyperfine components. Same contour levels as in the left panel
of Fig.~\ref{I05345_NH311_int} are shown.
The synthesised beam of the \AMM\ (1,1) 
VLA observations is shown in the bottom-left corner. The symbols have the same
meaning as in Fig.~\ref{I05345_NH311_int};
(b) same as panel (a) for the second-order moment.}
\label{moments}
\end{figure*}

\begin{figure*}
\centerline{\includegraphics[angle=-90,width=16cm]{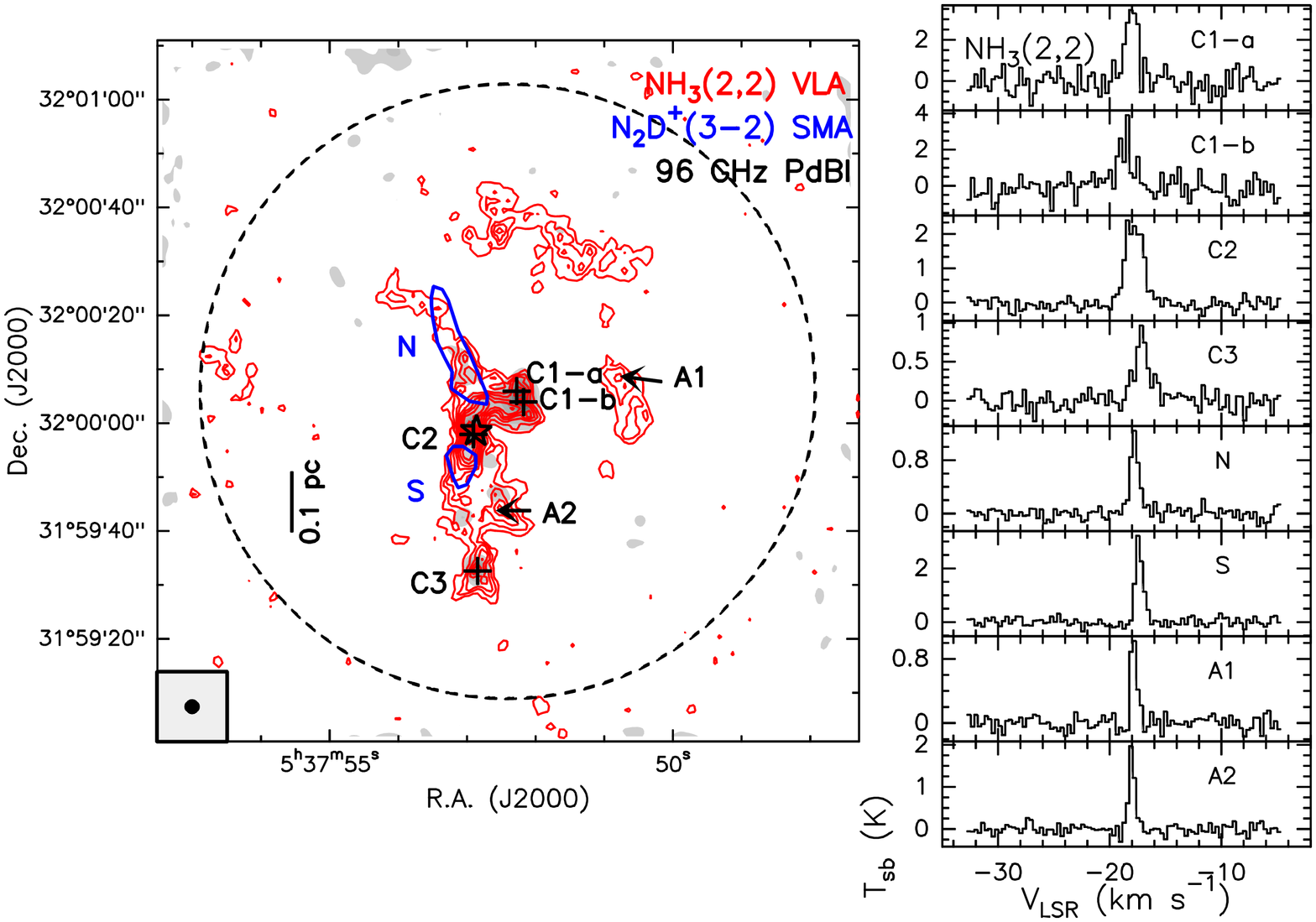}}
\caption{Same as Fig.~\ref{I05345_NH311_int} for the \AMM\ (2,2) transitions. 
In left panel, the \AMM\ (2,2) line emission has been integrated over the velocity interval
--19.02 and --16.55 \kms . Contour levels start from the 6$\%$ of the maximum
($\sim 0.015$ Jy beam$^{-1}$ \kms ), and are in steps of 10$\%$.}
\label{I05345_NH322_int}
\end{figure*}

\subsection{Integrated intensity maps, moments and averaged spectra of \AMM\ (1,1) and (2,2)}
\label{distribution}

In the left panel of Fig.~\ref{I05345_NH311_int} we show the 
zero-th order moment map of \AMM\ (1,1) obtained towards \i , by integrating
over the main group of hyperfine components.
The shape is irregular with a main central filamentary structure extended 
$\geq 80$\arcsec\ in the NS direction, and a second clumpy filamentary-like 
structure to the NW having an arch-like shape (significance level from 6 to 9$\sigma$), 
oriented roughly in the NE-SW direction.
%An isolated clump is marginally detected at the eastern edge of the primary
%beam. This clump, however, does not seem to be related to the
%central protocluster and will be not discussed in the following.}
The \AMM\ (1,1) integrated emission traces an envelope that encompasses
all the sources previously detected in the millimetre continuum and \D\
(Fig.~\ref{I05345_NH311_int}, we adopt the same nomenclature
as in  paper~II), in agreement with the relatively low critical
density of the transition ($\sim 10^{4}$ \cmc ). The largest integrated intensity is
detected towards the millimetre continuum cores, i.e. in the direction of active 
star formation (see Sect.~\ref{Introduction}).
%especially towards C2 which likely harbours
%the most evolved and most massive object of the protocluster.

Although it is difficult to identify separate cores given the complex 
emission shape, we point out two additional ammonia peaks 
labelled as A1 and A2, seen also in \H\ (see Sect.~\ref{comp_maps}).
Both peaks are undetected in the infrared, as we will show
in Sect.~\ref{infraredimages}.
%A1 is the most prominent clump in the "arch-like" filament to the NW,
A2 is in between the continuum sources C1 and C3 in the 
central filament and located close to a marginal detection in the 
millimetre continuum (Fig.~\ref{I05345_NH311_int})
called C4 in paper~II. C4 is a marginal detection in the millimetre continuum, 
while A2 is a well defined ammonia condensation, and it does not peak 
exactly at the position of C4. Therefore we cannot be sure that
the two cores are associated. Note that we did not include C4 
in any of the figures to make them clearer.

Fig.~\ref{moments} shows the maps of the first- and second-order moments 
of the main group of hyperfine components of the (1,1) line.
The first-order moment (panel a) does not show very steep velocity
gradients. 
%Red-shifted emission is noticeable in the arch-like filament
%north of N, and in the central region of the filament in the east-west direction, 
%roughly in between C2 and S.
%This velocity perturbation could be due to an ionised
%jet or an UC \HII\ region, possibly associated with C2, pushing material away. }
%Blue-shifted emission is noticeable across both the arch-like
%filament, south of C1-b and north-west of A2.
%In all cases, this can be caused by the presence of the blue lobe of 
%a widespread outflow that will be presented in Sect.~\ref{dense}.
The second-order moment (panel b) clearly shows 
an enhancement of the velocity dispersion towards the continuum cores, 
i.e. close to active star formation.
There is also evidence of line broadening towards the arch-like
NW filament, around 30\arcsec\ north of the continuum source C1, 
with values up to 1 \kms\ (corresponding to line widths
of 2.35 \kms ).
Both the velocity field and the velocity dispersion across I05345 will be 
examined more extensively in Sect.~\ref{sect_maps}.

The \AMM\ (2,2) zero-th order moment map is shown in the left panel of
Fig.~\ref{I05345_NH322_int}. The emission is more 
compact than that of the (1,1) line, and arises mostly from the mm-continuum cores. 
Significant integrated emission is detected towards the starless cores
S, N, A1 and A2. The arch-like filament to the NE is also clearly detected,
but appears less extended than in the \AMM\ (1,1) map.
For this line we do not show the first and second order moment maps which
do not give any additional information with respect to that presented in Fig.~\ref{moments}.

In the right panels of Figs.~\ref{I05345_NH311_int} and ~\ref{I05345_NH322_int}
we show the spectra of \AMM\ (1,1) and (2,2), respectively, averaged over all
identified cores. The contours used to extract the spectra are: the 3$\sigma$ rms level
of the 284~GHz continuum for C1-a and C1-b (paper ~II); the 3$\sigma$ rms level
of the 225~GHz continuum for C2 and C3 (paper~I); the 3$\sigma$ rms level
of the \D\ (3--2) integrated emission for N and S (paper~I).
We used the 3$\sigma$ rms level of the \AMM\ (2,2) emission to extract the
spectra of A1 and A2. 
Both lines are systematically broader towards the continuum cores, showing 
once again that the other cores are more quiescent. 
%We note that towards A1 there is almost no 
%emission in the integrated map (left panel in Fig.~\ref{I05345_NH322_int})
%but the \AMM\ (2,2) spectrum shows a clear and strong detection (right panel
%in Fig.~\ref{I05345_NH322_int}). 
%This is due to the fact that the emission is so narrow towards A1 
%that the intensity is diluted in the broader velocity range adopted to create the map.

The (1,1) lines were fitted by taking into account their hyperfine structure 
using METHOD NH3 of the CLASS package\footnote{This method assumes that 
all the hyperfine components have the same excitation temperature and width, and 
that their separation is fixed to the laboratory value.
The method also provides an estimate of the optical depth of 
the line, based on the intensity ratio of the different hyperfine components.
See the CLASS manual at http://iram.fr/IRAMFR/GILDAS/doc/html/class-html/class.html for details.}.
For the (2,2) transition, the hyperfine fitting procedure did not
give good results because the magnetic hyperfine components
of the main central line were not resolved (expected maximum separation of
$\sim 0.12$ \kms ), and the inner satellites of the electric structure
were not included in our observed bandwidth, so that we decided to fit the 
lines with single Gaussians.
% We stress, however, that this assumption should not affect significantly 
%the parameters derived, because for the lines with good hyperfine fits to the
%(2,2) line \Tr\ differ from that estimated assuming a Gaussian line by less than 
%15\%, which means that our method introduces a correction comparable to the 
%uncertainty on the flux calibration.}

The results are listed in Table~\ref{line_par}. For each core, small
differences are noticeable between $V_{\rm peak}$ of the (1,1) and
(2,2) lines beyond the indicated error margins
(e.g., for C1, C1-b, and N), probably because of the different
fit method that does not take into account the hyperfine structure
for the (2,2) line (see above). This could also explain the fact that 
$\Delta v$ is found to be slightly but systematically smaller for the 
(1,1) than for the (2,2) line. However, in the protostellar cores this 
can have another origin: because the (2,2) transition needs higher
temperatures to be excited, it likely arises from gas
closer to the forming protostars (as the integrated intensity
maps in Figs.~\ref{I05345_NH311_int} and \ref{I05345_NH322_int} suggest), 
and hence likely more turbulent. 
% cambiamento precedente per rispondere all'editore
%Small differences (of the order of $\sim 1$ \kms ) in the peak
%velocities among the cores can also be noticed.

\begin{table}
\begin{center}
\caption[]{Derived line parameters for the cores from the spectra integrated
above the 3$\sigma$ level. 
The uncertainties obtained from the fitting procedure are in parentheses.}
\label{line_par}
\begin{tabular}{ccccc}
\hline \hline
\multicolumn{5}{c}{ \AMM\ (1,1)  \tablefootmark{a}  } \\
core &  $A \times \tau_{m}$ & $V_{\rm peak}$     & $\Delta v$ & $\tau_{m}$  \\
  & (K)        & (\kms ) &  (\kms )                        & \\
\hline 
C1 &  3.3(0.6) & --18.24(0.02) &  1.42(0.05) &     0.4(0.2)  \\
C1-a &  8.5(1) &	--17.97(0.02) &      0.94(0.07) &	 0.5(0.3) \\ 
C1-b &  5.2(0.9) 	&      --18.40(0.04) &      1.3(0.1) &	 0.3(0.4) \\ 
%C2     & 73(2) &	--17.822(0.009) &      1.48(0.02) &	 1.9(0.1) \\ 
C2     &  11.7(0.2) &	--17.822(0.009) &      1.48(0.02) &	 1.9(0.1) \\ 
C3     &  3.7(0.3)	 & --17.1(0.02)    &  1.06(0.05)	& 1.5(0.3) \\ 
N       &   10.1(0.9)	 & --17.829(0.005) &      0.63(0.02) &	 1.7(0.15) \\ 
S       &   19.2(0.6)   & --17.438(0.007)  &    0.66(0.02)  &	 2.8(0.2) \\ 
A1    &   4.4(0.6)  & --17.900(0.007) &  0.63(0.02) & 0.5(0.2) \\
A2    &  12.1(0.7) & --18.00(0.01) &  0.58(0.02) &  1.75(0.2) \\
\hline
 & & & & \\
 \multicolumn{5}{c}{ \AMM\ (2,2) \tablefootmark{b} } \\
core  & Area & $V_{\rm peak}$    & $\Delta v$ & $T_{\rm sb}^{\rm peak}$  \\
  & (K \kms )        & (\kms ) &  (\kms )   & (K)            \\
\hline
C1 &  2.8(0.1) & --18.37(0.04) &    1.95(0.1) &    1.34(0.04) \\
C1-a &  5.0(0.5) &     --18.04(0.06) &      1.25(0.14) &	  3.7(0.1) \\
C1-b &  4.6(0.6) &     --18.6(0.1) &	    1.7(0.3) &	  2.5(0.1) \\ 
C2 &  4.8(0.2) &    --17.81(0.03) &      1.87(0.07) &	  2.44(0.03) \\
C3 &  1.3(0.1) &     --17.13(0.06) &      1.5(0.2)  &	  0.80(0.02) \\ 
N &  1.1(0.1) &    --17.78(0.02) &     0.87(0.06) &	  1.20(0.04) \\ 
S &  2.57(0.07)   &  --17.38(0.02) &     0.78(0.04) &	  3.11(0.03) \\ 
A1 &  0.7(0.1)  &  --17.86(0.02) &    0.61(0.05) &   1.05(0.03) \\
A2 &  1.27(0.09) & --18.02(0.02) &   0.60(0.04) &   1.78(0.03) \\
\hline 
\end{tabular}
\end{center}
\tablefoot{
\tablefoottext{a}{Cols. 2--5 list the results of the CLASS hyperfine fitting
procedure: $A \times \tau_{m}$ = $f[J_{\nu}(T_{\rm ex})-J_{\nu}(T_{\rm BG})]$, where
$f$ is the filling factor (assumed to be unity), $J_{\nu}(T_{\rm ex})$ and $J_{\nu}(T_{\rm BG})$
are the equivalent Rayleigh-Jeans excitation and background temperatures, respectively,
$\tau_{m}$ is the opacity of the main group of hyperfine components;
$V_{\rm peak}$ = peak velocity; $\Delta v$ = full width at half maximum
corrected for hyperfine splitting; $\tau_{m}$ = opacity of the main group of hyperfine components;}
\tablefoottext{b}{Cols. 2--5 give the results of Gaussian fits: Area = total
integrated intensity; 
$V_{\rm peak}$ = peak velocity; $\Delta v$ = full width at half maximum;
$T_{\rm sb}^{\rm peak}$ = peak intensity in synthesised beam temperature units.}
}
\end{table}

%The averaged maps are superimposed
%on the 3~mm continuum emission observed by paper~I with the
%PdBI with an angular resolution of $\sim 3$\arcsec .
\subsection{Comparison with other tracers}
\label{comp_maps}

\begin{figure}
\centerline{\includegraphics[angle=-90,width=8.5cm]{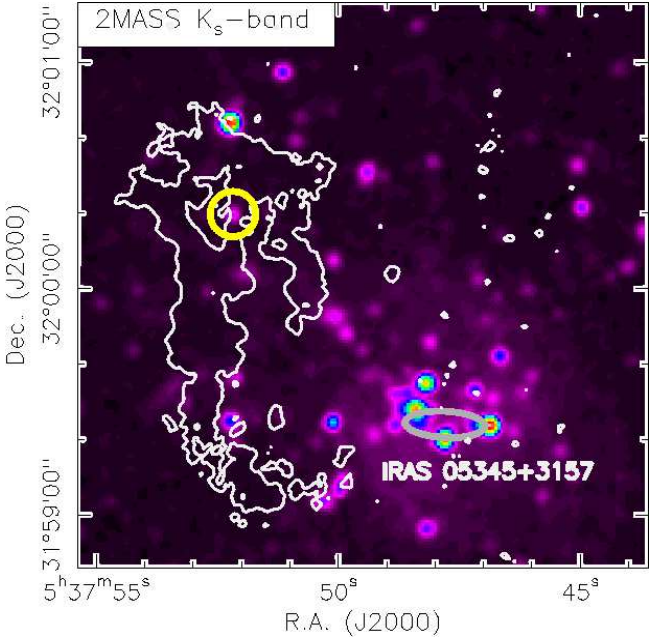}}
\caption{2MASS $K_s$-band image of the region associated with 
IRAS 05345+3157 (colour scale). The white contour represents the 6$\%$ level of
the maximum ($\sim 0.039$ Jy beam$^{-1}$ \kms ) of
the \AMM\ (1,1) zero-th order moment map.  The yellow circle 
centred roughly at Dec. +32$^{\circ}$00$^{\prime}$20$^{\prime \prime }$ 
pinpoints the location of the 4.5 $\mu$m source discussed
in the text (Sect.~\ref{infraredimages}). The position error 
of the IRAS source is indicated by the grey ellipse. }
%\newline
%Right panel: zoom of the image in the left panel in a 80\arcsec\ square region 
%centred around the phase centre of the VLA ammonia map. 
%%The white contours represent the \AMM\ (1,1) integrated
%%emission (contour levels are 0.0012, 0.0048 and 0.0084 Jy beam$^{-1}$ \kms , respectively). 
%The VLA synthesised beam is shown in the bottom-left corner. 
% The other symbols indicate the cores as identified in Fig.~1: crosses = emission
%peaks of C1-a, C1-b, C2, C3, A1 and A2; contours = 3 $\sigma$ rms level
%of the \D\ (3--2) emission identifying N and S; star = 3.6~cm continuum
%emission peak.}}
\label{2mass}
\end{figure}

\subsubsection{Near- and mid-infrared continuum images}
\label{infraredimages}

\begin{figure}[!]
{\includegraphics[angle=-90,width=8.8cm]{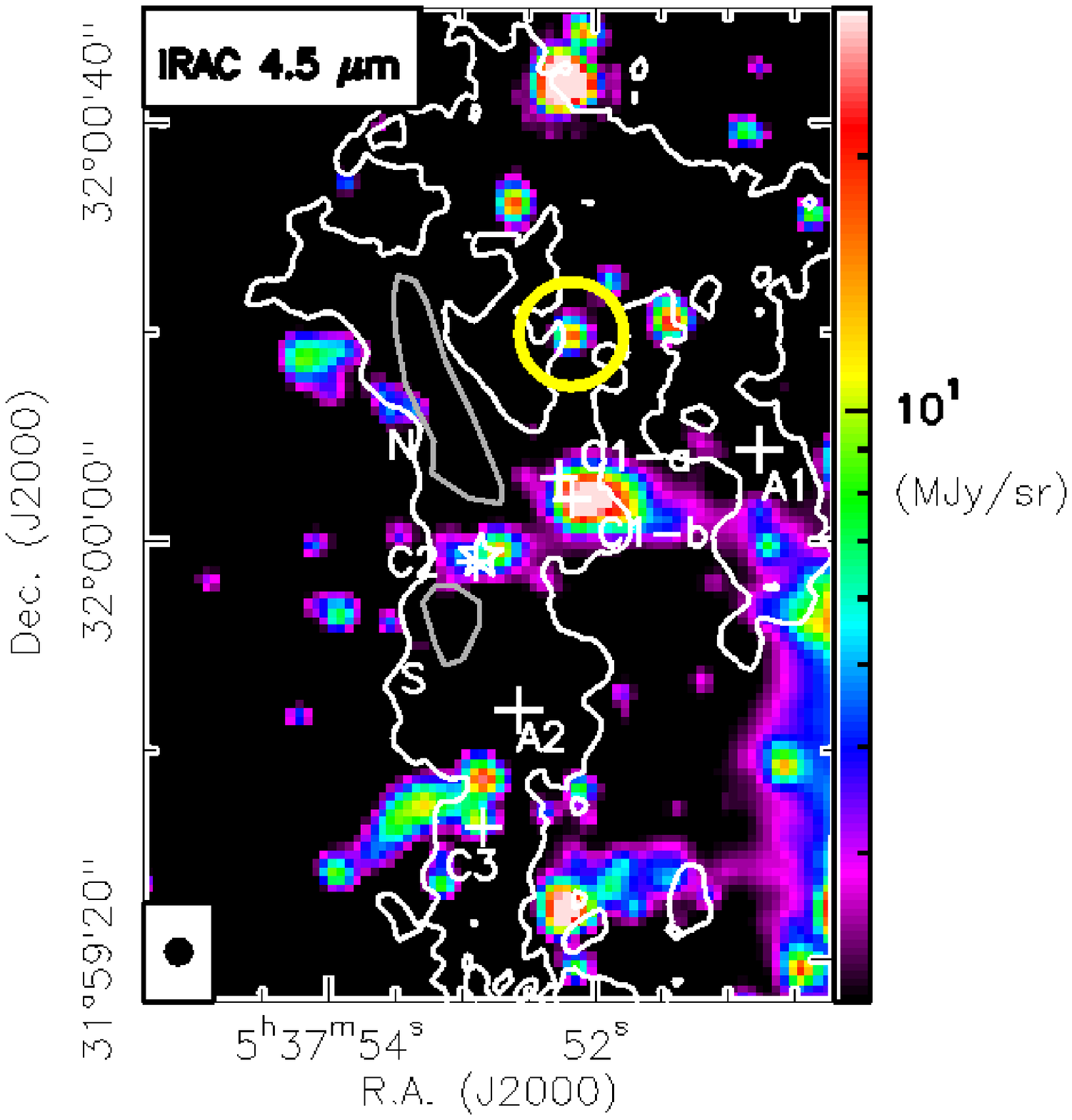}}
{\includegraphics[angle=-90,width=8.8cm]{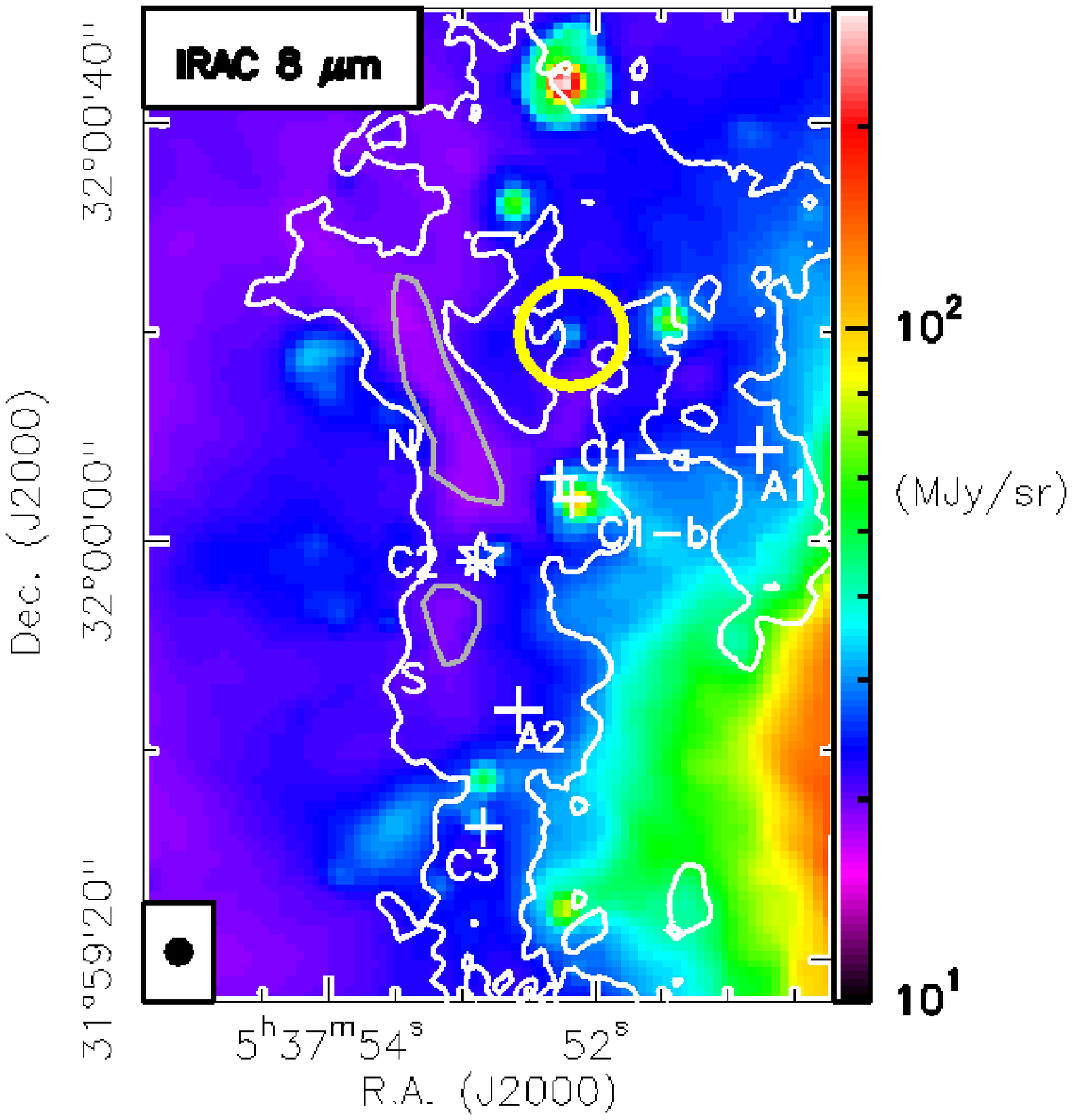}}
\caption{Mid-infrared emission observed towards I05345 with the Spitzer 
Space Telescope. The two colour images show the Spitzer IRAC bands centred at
4.5 (top panel) and 8 $\mu$m (bottom panel). The white contour
indicates the ammonia (1,1) emission and is the same as in Fig.~\ref{2mass}. 
 The yellow circle about 15\arcsec\
above C1-a pinpoints the location of the 4.5 $\mu$m source discussed
in the text (as in Fig.~\ref{2mass}), and the other symbols indicate the 
location of the gaseous cores as in Fig.~\ref{I05345_NH311_int}. The two grey
contours represent the 3$\sigma$ level of the \D\ (3--2) integrated
emission, and are the same as the blue contours in Fig.~\ref{I05345_NH311_int}.
 In each panel, The VLA synthesised beam is
shown in the bottom-left corner.
}
\label{irac}
\end{figure}

The two ammonia filaments shown in Fig.~\ref{I05345_NH311_int}
are located about $\sim 1$\arcmin\ north-east of the near-infrared 
cluster corresponding to the IRAS source, as it can be seen from
the 2MASS ($K_s$-band) image (Fig.~\ref{2mass}). 
Two of the four IRAC images (band central wavelengths 4.5 and 8 $\mu$m)
of the region, taken with the Spitzer Space Telescope, are shown in
Fig.~\ref{irac}. We do not show the 3.6 and 5.8 $\mu$m IRAC images because
they are very similar to the 4.5 $\mu$m one. The star-forming cores C1-a, C1-b, C2 and C3 
are all associated with mid-infrared emission, while the starless cores
N, S, A1 and A2 lie in
infrared-dark regions (see Figs.~\ref{2mass} and \ref{irac}).
%Specifically,
%N and S are detected  towards two of the mostly mid-infrared dark
%regions}, especially apparent in the 8 $\mu$m image.
%The 4.5 $\mu$m emission detected towards C1-b is elongated to the west
%(top panel in Fig.~\ref{irac}), 
%due either to a companion or to warm gas associated with an outflow. }
%%The latter possibility seems to us the most likely
%%one,  because a widespread outflow detected in \CO\ is
%%located in that region (see paper~II, see also Sect.~\ref{dense}).} 
%The emission towards C2
%also looks elongated east-west in the 4.5 $\mu$m image, and
%resembles the red-shifted emission noticed 
%in the ammonia first order moment map (left panel in Fig.~\ref{moments})
%across C2,  suggesting that the two phenomena could
%be related.} 
% We highlight the presence of a bright infrared source 
%located few arcseconds north of core C3.
%Some extended emission is noticeable close to A1 in the 
%4.5 $\mu$m band, probably
%contaminated by the IRAS source emission. 
In both the 2MASS and 
4.5 $\mu$m image, we highlight the presence of a point-like source located approximately
15\arcsec\ north of C1-a  (see yellow circle in Figs.~\ref{2mass} and \ref{irac}), 
in between C1-a and the north-western
filament, which seems placed at the centre of a region almost
devoided of ammonia emission. We will discuss better this
source and its possible influence in the surrounding gas in
Sect.~\ref{velocity}.
The nature of the (near- to far-)infrared emission in I05345
will be discussed through deeper sensitivity and angular resolution images
in a forthcoming paper (Fontani et al. in prep.).

\subsubsection{High density gas and outflow tracers}
\label{dense}

\begin{figure*}
\centerline{\includegraphics[angle=-90,width=6.8cm]{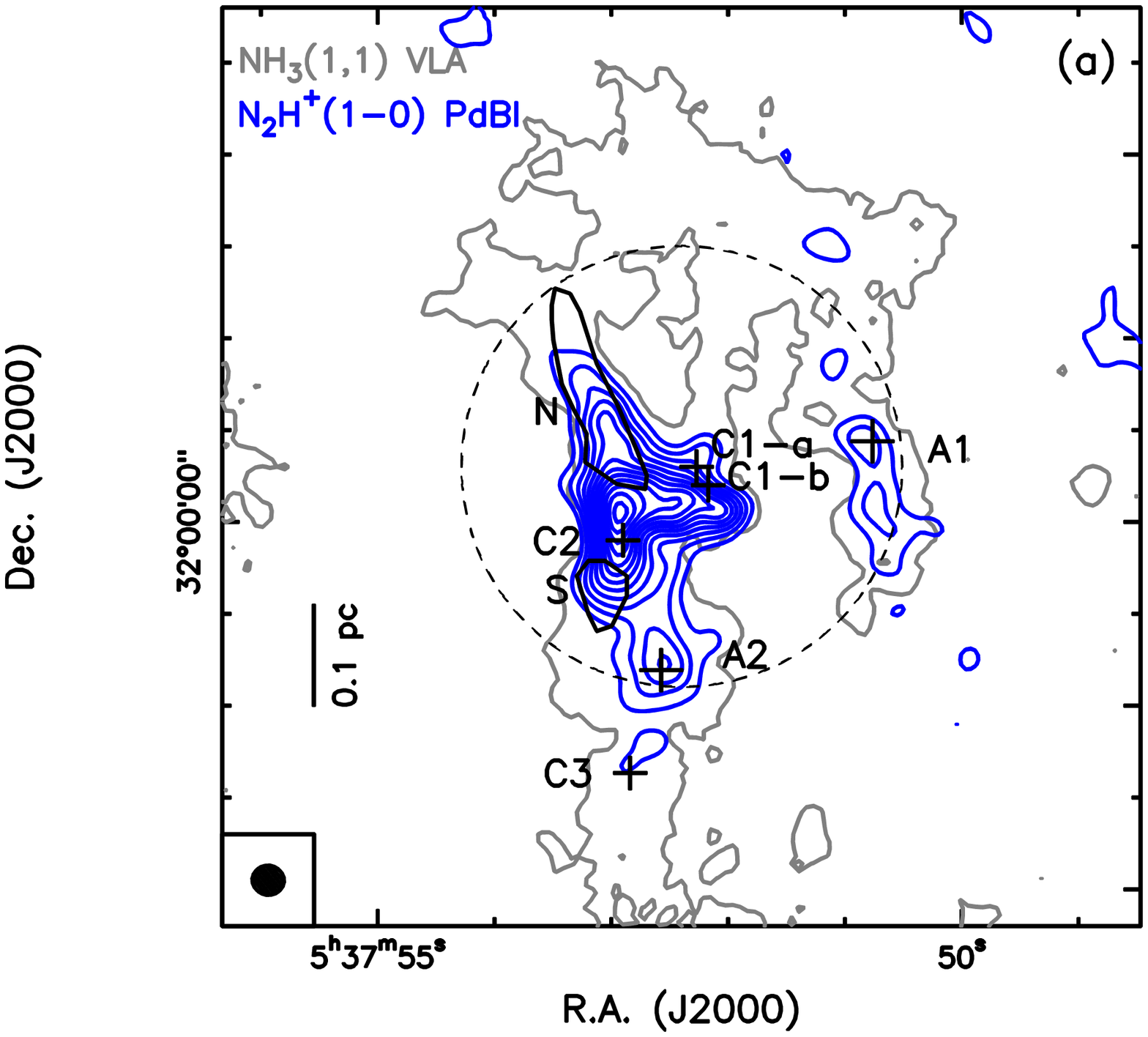}
                    \includegraphics[angle=-90,width=5.5cm]{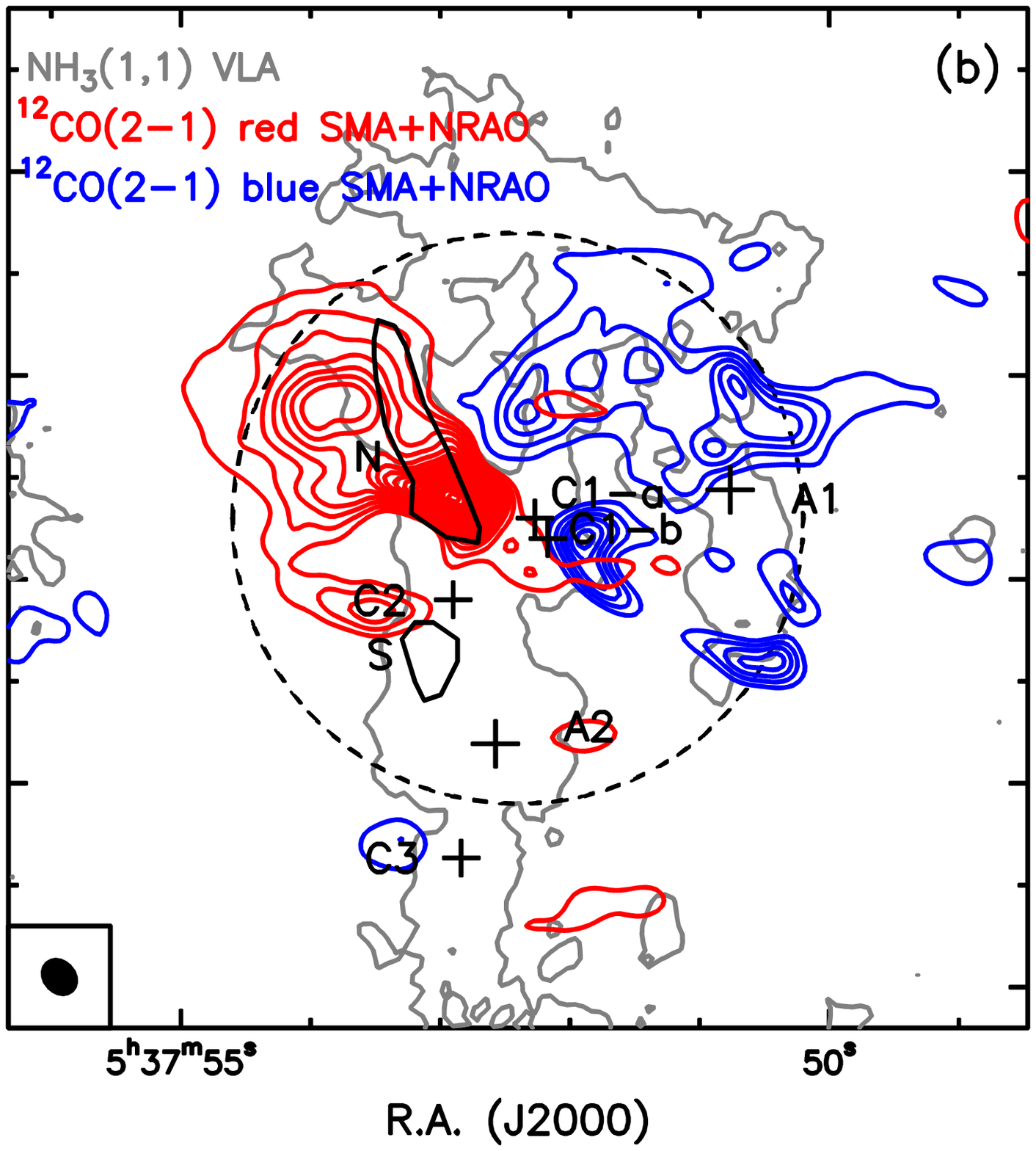}
                    \includegraphics[angle=-90,width=5.5cm]{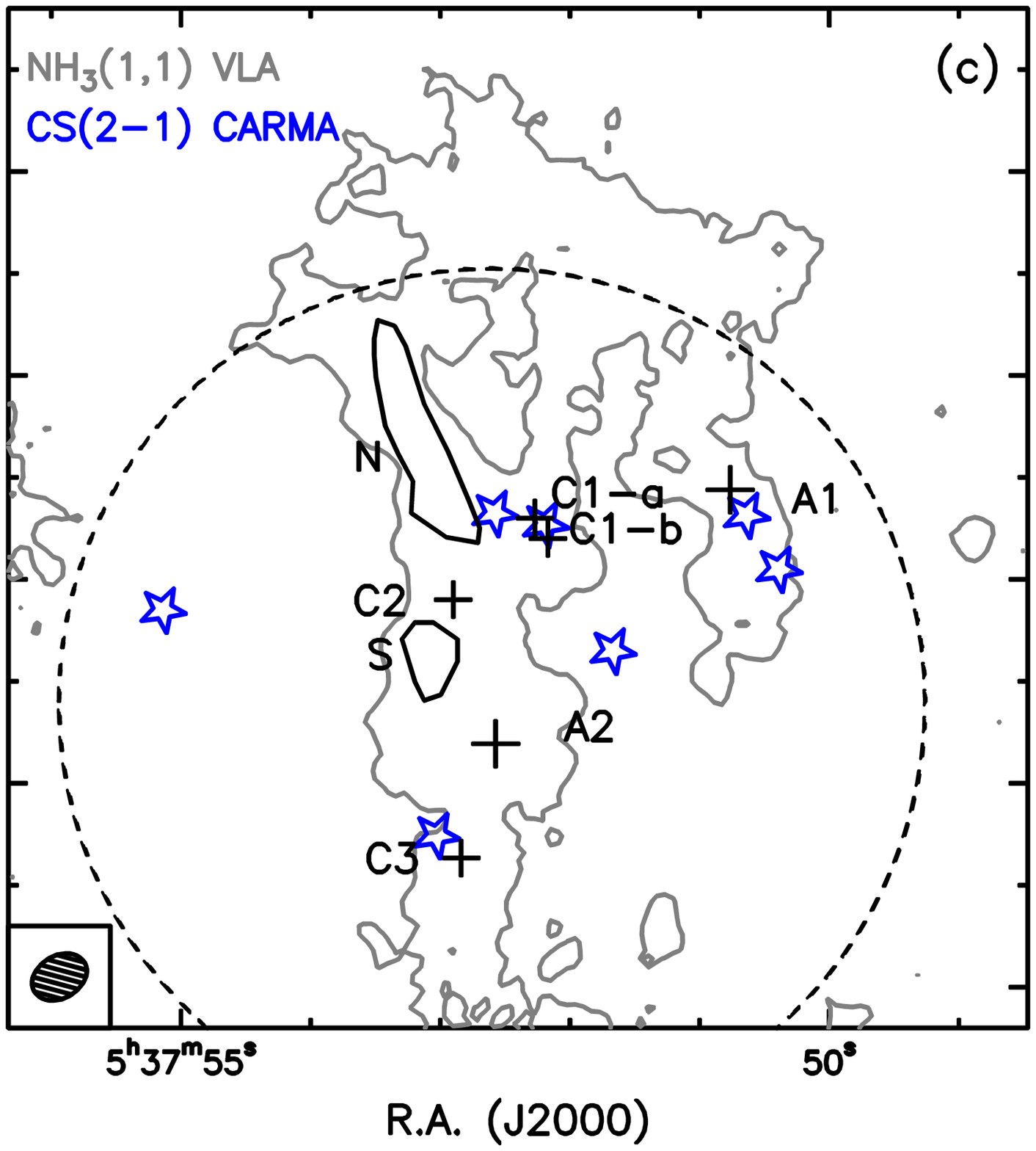}}
\caption{Overlay between the \AMM\ (1,1) integrated emission (grey
contour, same level as in Figs.~\ref{2mass} and \ref{irac}), and the
integrated emission of three previously observed tracers. They are, from
left to right: (a) the main group of hyperfine components of \H\ (1--0)
observed with the PdBI (paper~I, blue contours. First level and step = 0.02 Jy beam$^{-1}$); 
(b) the \CO\ integrated emission over the red and blue non-Gaussian wings 
derived from SMA + NRAO merged data (same contours as in paper~II); (c) the CS (2--1) line
emission peaks detected with CARMA (Lee et al~\citeyear{lee}, peak
positions are taken from their Table~1).
\newline
In each panel, black contours indicate the 3$\sigma$ rms contour level of the 
\D\ (3--2) emission, and crosses mark the peak position of the 3~mm 
continuum sources and of the ammonia cores as in Figs.~\ref{I05345_NH311_int} 
and \ref{I05345_NH322_int}. The ellipse in
the bottom left corner and the dashed circle represent the synthesised beam 
and the primary beam, respectively, of the instrument
used to observe the tracer superimposed on the ammonia emission: PdBI (left panel); 
SMA + NRAO (middle panel); CARMA (right panel).}
\label{amm_n2hp}
\end{figure*}

The ammonia emission is in very good agreement with that 
of \H\ as can be seen in panel (a) of Fig.~\ref{amm_n2hp},
where we superimpose the integrated intensity of the main group of hyperfine
components of \H\ (1--0) (paper~I)
on the \AMM\ (1,1) emission map shown in
Fig.~\ref{I05345_NH311_int}. 
Because the primary beam of the \H\ map is smaller that that of the \AMM\ maps,
the \H\ emission looks less extended, but inside the PdBI primary beam 
($\sim 48$\arcsec ) the two tracers are almost coincident,
although \H\ (1--0) is expected to trace gas at slightly higher densities
($n_{\rm crit} \sim 10^{5}$ \cmc ).
\H\ (1--0) is clearly detected towards both A1 and A2 indicating that
these are real physical structures, even though in paper~II we did not
discuss them because they fall at the edge of the PdBI primary beam.
A more exhaustive comparison between 
\AMM\ and \H\ will be presented in Sect.~\ref{abundances}.

In panel (b) of Fig.~\ref{amm_n2hp} we plot the overlay between the 
\AMM\ (1,1) emission and the widespread outflow detected in 
\CO\ and reported in paper~II.
 The internal part of the blue lobe of the outflow coincides spatially
with the elongated mid-infrared emission detected west of C1-b (compare
panel (b) of Fig.~\ref{amm_n2hp}
with Fig.~\ref{irac}), suggesting that this emission could arise from the
warm gas associated with the outflow.
%The ammonia "arch-like" filament NW of the field centre spatially 
%coincides (in projection) with part of the widespread blue lobe of the
%molecular outflow. 
%%Because ammonia is thought to be also an
%%outflow tracer, part of the \AMM\ emission detected in this position
%%could come from the outflow.
%However, because the velocity of the ammonia lines is
%different from that of the gas in the outflow, probably
%the two emissions are not associated.
%tolto il discorso della cavity, come sopra

CARMA CS (2--1) observations of the region (Lee et al.~\citeyear{lee})
reveal seven emission peaks, which are plotted in panel (c) of 
Fig.~\ref{amm_n2hp}. CS emission is detected
towards C1, C3, A1, and close to the southern part of N.
%Two additional peaks are found several arcseconds to the east and
%west of core C2. 
% None of the CS emission peaks coincides
%with an emission peak of \AMM\ (1,1) and/or (2,2), but this can
%be due either to the different critical densities or to optical
%depth effects.}
% The emission peak close to N coincides in projection with 
%part of the red lobe of the CO outflow (middle panel).}
Cores C2, A2, S and N are not detected in CS.
Because CS is well known to be highly depleted towards low-mass pre--stellar
cores (e.g.~Tafalla et al.~\citeyear{tafalla02}), the non detection
in cores N and S could be due to molecular freeze-out, which would 
support the conclusions of paper~I and II that N and S are likely 
in the pre--stellar phase.
Core A2, not detected in CS, marginally detected in the millimetre
continuum (C4 in  paper~II), and clearly detected only in the non-depleted 
species \AMM\ and \H , might be another very young protocluster member
where CS is depleted like in the pre--stellar core candidates.
On the other hand, the non-detection of core C2
is puzzling since this condensation likely harbours the most evolved
(and maybe most massive) object of the protocluster. 

\subsection{Derivation of physical parameters from ammonia}
\label{derivation}

\subsubsection{Rotation temperature and \AMM\ column density}
\label{temperature}

From the averaged ammonia spectra shown in Figs.~\ref{I05345_NH311_int} and \ref{I05345_NH322_int},
fitted as explained in Sect.~\ref{distribution}, we have derived several physical parameters.
Rotation temperatures, \Tr , and column densities, $N$(\AMM ), were calculated from 
the (2,2)-to-(1,1) line intensity ratio as described in Appendix A of Busquet et al.~(\citeyear{busquet09}). 
The method, based on Ho \& Townes (1983), assumes that: (1) the inversion doublets constitute
a two-level system; (2) the excitation temperature, \Tex , and line
widths at half maximum, \deltav , are the same for both transitions; (3) the physical conditions
of the gas are homogeneous along the line-of-sight, so that if gradients are
present, the derived parameters are average values along the line-of-sight.
In the derivation of the parameters, we also assumed a unity filling factor.
We stress that the \Tr\ derived using the results of Gaussian fits for the 
(2,2) transition (see Sect.~3.1) should differ only of 15$\%$ from the \Tr\ that 
we would derive by fitting the hyperfine structure.

\Tr\ and $N$(\AMM ) are given in Table~\ref{temp}.
As expected, the coldest condensations are the pre--stellar core candidates N and S,
with a \Tr\ of 13 and $\sim 16$ K, respectively, while the continuum cores C1-a, C1-b,
C2 and C3 have temperatures larger than 17~K. 
Because C1-b harbours a hot-core (detected in paper~II through CH$_3$CN and OCS lines), 
the rotation temperature derived for this condensation is representative only of the 
lower-density envelope in which the hot-core is embedded.
 The source averaged ammonia column densities
are different among the cores, going from $\sim 4 \times 10^{14}$ \cmq\ 
for C1-b and C3 to $\sim 1.3 \times 10^{15}$ \cmq\ for C2. 
We will come back over this point in Sect.~\ref{abundances}.
We stress that the uncertainties on the column densities given in Table~\ref{temp}, 
which are very small (less than 10\%), do not take into account the error in the 
flux calibration, typically of about $10-15$\%. 
Therefore, an uncertainty of about 30\% seems more realistic for
$N$(\AMM ).

\subsubsection{Core gas masses}
\label{masses}

%From the line widths at half maximum given in Table~\ref{line_par}, we have 
%computed the masses of the cores assuming virial equilibrium and negligible
%contributions of magnetic field and surface pressure. 
%Assuming also that the cores are spherical and homogeneous, 
%one can demonstrate that the gas mass is given by (see Eq. (3) in 
%MacLaren et al.~\citeyear{maclaren}):
%\begin{equation}
%M_{\rm VIR}(M_{\odot})\simeq 210 \Delta v^{2} {\rm (\kms )}\,R ({\rm pc})\;\;, 
%\end{equation}
%where $R$ is the core radius derived from interferometric
%observations (paper~I,  paper~II).
%We adopted the line widths derived from the (1,1) transition which
%take into account the hyperfine structure.
%The virial masses are in good agreement with the values given
%in paper~II based on the \H\ line widths.

The gas mass can be estimated from the ammonia total column density 
from the relation:
\begin{equation}
M_{\rm N(NH_3)}=\frac{\pi}{4 \ln{2}} D^{2} \mu m_{\rm H} \frac{N(\rm{NH_3})}{X} \;\;, 
\end{equation}
where $\mu=2.33$ is the mean molecular weight, $m_{\rm H}$ is the hydrogen mass, 
$D$ is the linear diameter, $N(\rm{NH_3})$ is the 
ammonia total column density and $X$ is the [\AMM /H$_2$] abundance ratio. 
We adopt a uniform density as zero-th order assumption 
(i.e. $N{\rm (NH_3)}=n{\rm (NH_3)}\times D$) 
because we do not know the density profile of the condensations.
%We adopted $X=3 \times 10^{-8}$ as in Wu et al.~(\citeyear{wu}).

To have an estimate of the \AMM\ abundance in our region,
we can use the H$_2$ column densities computed in paper~II
from the millimetre continuum emission (see Table 3 of paper~II): 
we recomputed the masses from the optically
thin millimeter continuum emission using Eq. (1) of paper~II,
assuming the temperatures derived from \AMM\ in this work (and
a gas-to-dust ratio of 100). For C1-b, which harbours a hot-core,
we have used both the ammonia temperature,
representative of the lower-density envelope of the core, and that computed
from the hot-core tracer CH$_3$CN in paper~II (i.e. $\sim 200$ K).
This implicitely assumes coupling between gas and dust, which seems
reasonable at the densities of our cores ($\geq 10^{6}$ \cmc ). 
We derive $X\simeq 1\times10^{-8}$
and  $X\simeq 0.85\times10^{-8}$ for cores C1 and C2, respectively. 
In massive star-forming regions $X$ is found to be of the order of 10$^{-7}$ to  
10$^{-8}$ (Harju et al.~\citeyear{harju}, Wu et al.~\citeyear{wu}).
We decided to adopt $X= 10^{-8}$.
As in paper~II, we do not consider the continuum sources close to or beyond 
the interferometer fields of view (namely C3 and C4).
The various mass estimates are listed in Table~\ref{temp}.
We point out that the dust masses derived from the millimetre
continuum are affected by
factors of uncertainty which are difficult to quantify, especially 
the dust mass opacity coefficient and the gas-to-dust ratio
(see e.g. Beuther et al.~\citeyear{beuther02}), 
expected to introduce factors of 2 or more in the uncertainty of the
dust mass, which hence also affect both the H$_2$ column density 
and the \AMM\ abundance.
Therefore, both the dust mass estimates given in Table~2 and the
derived ammonia abundances must be regarded
with caution.

\begin{table}
\begin{center}
\caption[]{Physical parameters from ammonia for the cores.}
\label{temp}
\scriptsize
\begin{tabular}{cccccc}
\hline \hline
% \multicolumn{5}{c}{ \AMM\ (1,1)  \tablefootmark{a}  } \\
core &  \Tr\ & $N$(\AMM )  & $D$\tablefootmark{a}    & $M_{\rm N(NH_3)}$ & $M_{\rm cont}$  \\
  & (K)        & ($\times 10^{14}$\cmq ) & (pc) & (\solm ) & (\solm ) \\
\hline
C1  & 20(1) &  4(0.5) & 0.06 &  2.5 & 12\tablefootmark{b}  \\
C1-a &  21(1) &  6.1(0.7) & 0.016  &  0.3 & 5.9\tablefootmark{c}  \\ 
C1-b &  22(1) &  4.6(0.1) & 0.018  &  0.3 & 7.8\tablefootmark{c}; 0.6\tablefootmark{d}  \\ 
C2     & 17.0(0.3) &  13(2) & 0.05  &  7 & 6\tablefootmark{b}  \\ 
C3     & 17(1) &  4(1) & 0.06 &  3 & -- \\ 
N       &   13.0(0.3) &   6(1) & 0.09  &  9.5\tablefootmark{e} & -- \\ 
S       &  15.6(0.5) &  10(2) & 0.05  &  5\tablefootmark{e} & -- \\ 
A1   &  17.1(0.5) &  2(0.3) & 0.09  &  4 & -- \\
A2   & 15.0(0.4) &  6(1) & 0.07  &  6 & -- \\
\hline
\end{tabular}
\end{center}
\tablefoot{
\tablefoottext{a}{Derived from interferometric observations of: millimetre continuum 
(C1, C1-a, C1-b and C2, paper~II), \D\ (N and S,
 paper~I), and ammonia (A1, A2 and C3 this work);}
\tablefoottext{b}{from the 225~GHz continuum integrated emission (paper~II), assuming that \Tr\ equals the dust temperature and
a gas-to-dust ratio of 100;}
\tablefoottext{c}{from the 284~GHz continuum integrated emission (paper~II), assuming that \Tr\ equals the dust temperature
and
a gas-to-dust ratio of 100;}
\tablefoottext{d}{from the 284~GHz continuum integrated emission, assuming a hot core temperature of ~200K
(paper~II) and
a gas-to-dust ratio of 100.}
\tablefoottext{e}{core not well-defined in the ammonia maps.}
\normalsize
%procedure: Excitation temperature minus Background 
%temperature times the opacity of the main group pf hyperfine components
%($T_{ant}\times \tau_{m}$), peak velocity ($V_{\rm peak}$),
%full width at half maximum ($\Delta V$) corrected for hyperfine splitting, 
%opacity of the main group of hyperfine components ($\tau_{m}$);}
%\tablefoottext{b}{Cols. 2--5 give the results of Gaussian fits: total
%area (Area), peak velocity ($V_{\rm peak}$), full width at half maximum ($\Delta V$),
%peak flux ($T_{\rm sb}^{\rm peak}$);}
}
\end{table}

\section{Maps of the physical parameters}
\label{sect_maps}

From the datacubes of both \AMM\ (1,1) and (2,2) we have extracted spectra
on a grid of spacing $1\farcs6\times1\farcs6$. These have been
fitted as outlined in Sect.~\ref{distribution}. From the fit results, we
have derived maps of I05345 in several physical parameters,
the discussion of which is the main subject of this section.
%: the velocity
%field, the rotation temperature, the \AMM\ total column density, the non-thermal
%contribution to the velocity disperions, the \H\ - to - \AMM\ column density ratio.
%The discussion of these results is the subject of this section.

\subsection{Velocity field}
\label{velocity}

%\subsubsection{Central filament}

\begin{figure}
\centerline{\includegraphics[angle=-90,width=4.7cm]{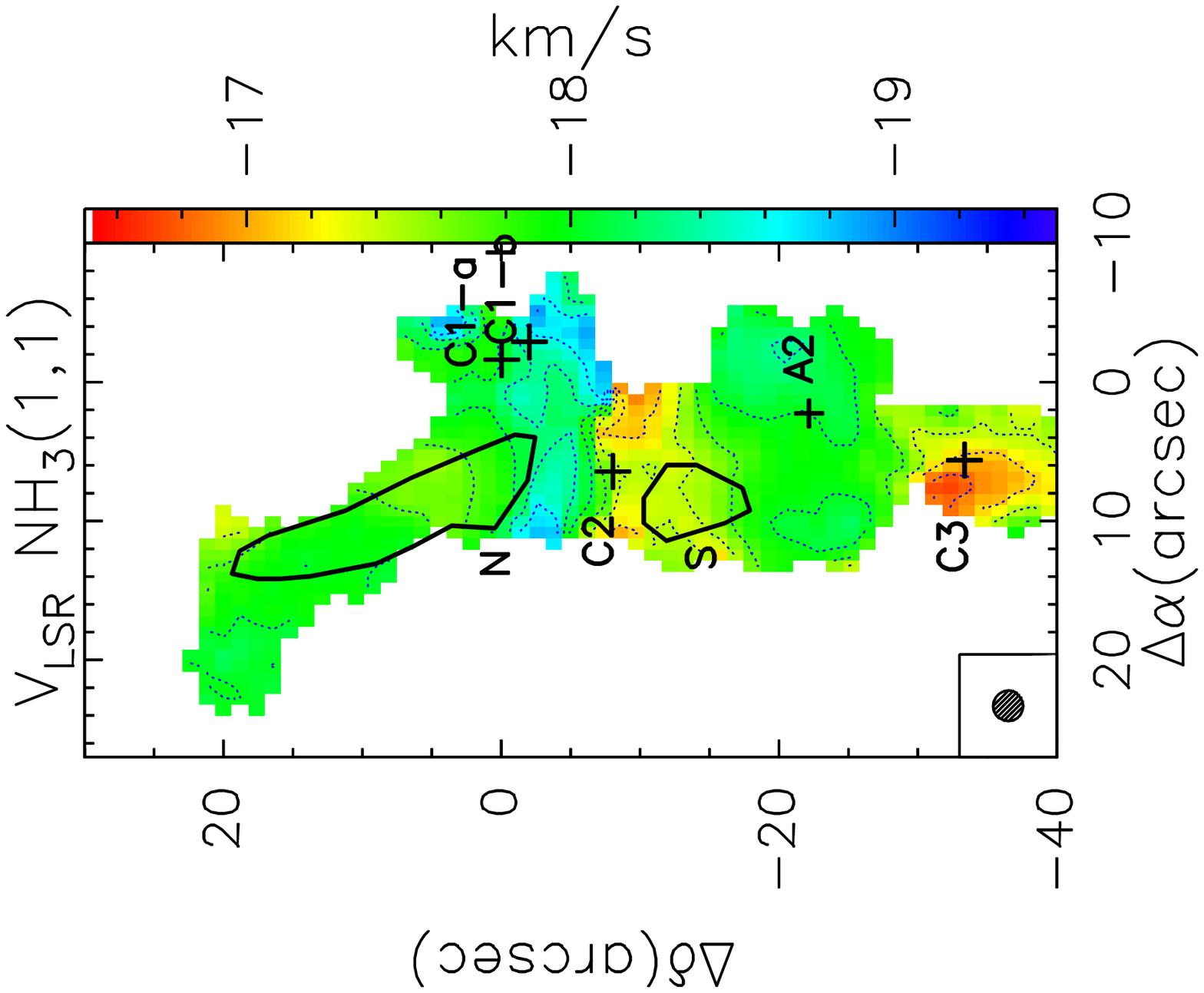}
                    \includegraphics[angle=-90,width=4.17cm]{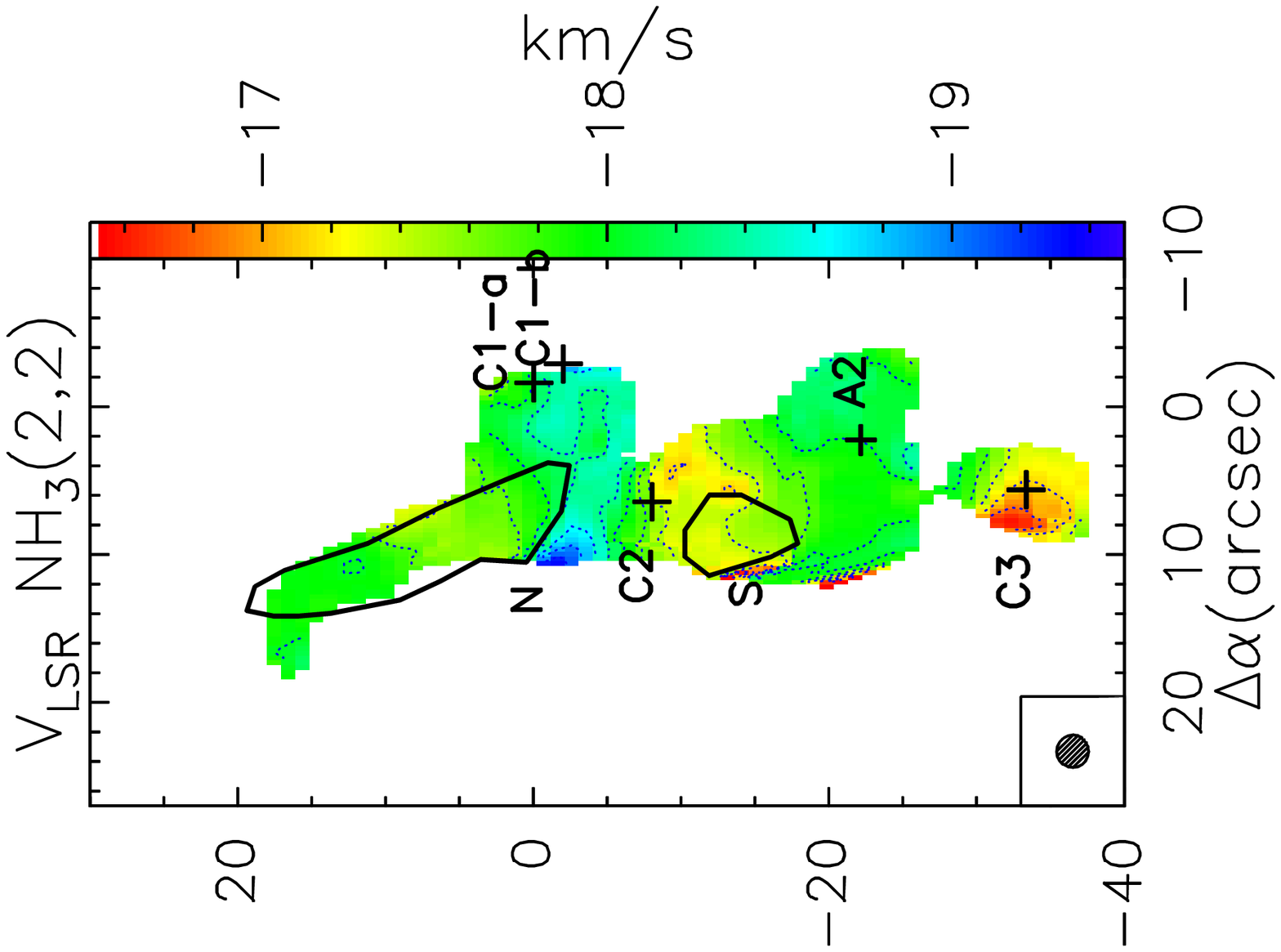}}
\caption{Map of the peak velocity of the \AMM\ (1,1) (left panel)
and (2,2) (right panel) inversion transitions fitted as outlined in Sect.~\ref{distribution}. 
Crosses indicate the emission peaks of cores C1-a, C1-b, C2, C3, and A2
(see Fig.~\ref{I05345_NH311_int}). Solid contours mark the location of the \D\ cores, 
(see Fig.~\ref{I05345_NH311_int}). In both panels, the dotted contours range
from --19.5 to --16.5 \kms , in steps of 0.3~\kms,
and the ellipse in the bottom left corner shows
the VLA synthesised beam.}
\label{velocity_maps}
\end{figure}

Maps of the peak velocity of \AMM\ (1,1) and (2,2) for the main filament 
are shown in Fig.~\ref{velocity_maps}. 
These maps are more accurate than that shown in panel a of Fig.~\ref{moments}
because they are based on the fitting method that takes into account the line
hyperfine structure. However, both maps agree quite well 
with the first-order moment map.
%We highlight again the evident red-shifted emission across C2 in the east-west direction,
%probably related to the mid-infrared emission seen in the 4.5 and 5.8 $\mu$m
%IRAC bands (Fig.~\ref{irac}). The gas could be blown by an ionised
%wind, possibly originating from C2, or swept away by an expanding \HII\ region.
%Perturbation due to the near-infrared cluster associated to
%the IRAS source is also possible (Figs.~\ref{2mass} and \ref{irac}).
%A velocity gradient towards core C3 is more apparent here than
%in Fig.~\ref{moments} from both the (1,1) and 
%the (2,2) peak velocities, and could be due either to outflow
%or to rotation. High-angular resolution observations centred on this
%source are needed to {shed light on} this issue. 
The maps in Fig.~\ref{velocity_maps} 
have been made interpolating the input data points on a regular grid. 
Because of this method, we recommend to regard with caution
regions containing several points with faint emission,
namely along the external edge of the emission 
and in the southern portion of the main filament.

\begin{figure*}
\centerline{\includegraphics[angle=-90,width=16cm]{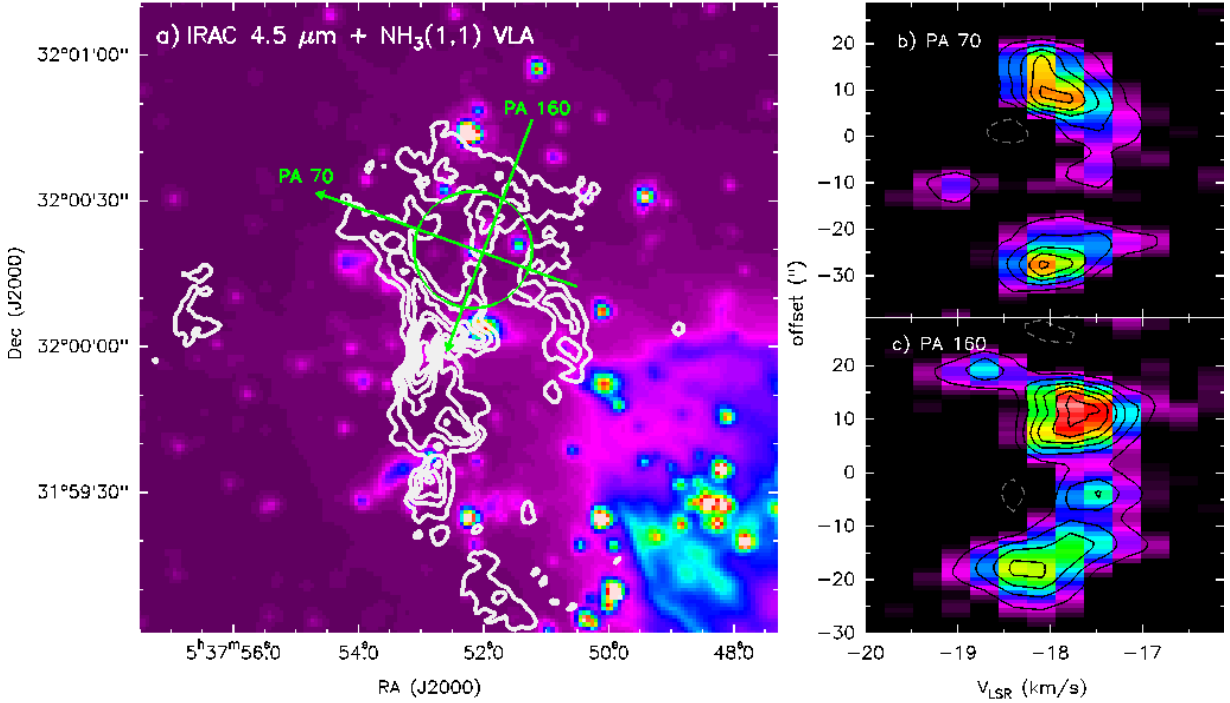}}
\caption{
 a) \AMM\ (1,1) integrated emission (white contours) overplotted on the 4.5 
$\mu$m Spitzer IRAC emission. Contours start at 10$\%$ of the peak intensity, 
(39.03 Jy beam$^{-1}$ m/s), and increase in steps of 15$\%$.
b) Position-velocity (p-v) plot of the \AMM\ (1,1) emission along the cut 
at PA 70\degr\ and centred on (--5\arcsec ,13\farcs 4) with respect to the phase 
centre, with positive offsets increasing as shown by the arrow in panel a).
c) idem as panel b) with PA 160\degr. Centre for both cuts is at 
position where the cuts intersect. In the p-v plots, contours start at 
--0.003 and increase in steps of 0.0015  Jy beam$^{-1}$.}
\label{pv}
\end{figure*}

In Sect.~\ref{infraredimages} we have remarked the presence of an infrared source
detected in both the 2MASS and the Spitzer-IRAC images
(see Figs.~\ref{2mass} and \ref{irac}), placed in between the main
filament and the arch-like northern one at the
centre of a region devoided of ammonia emission. Such a feature
resembles the 'hole' predicted by the model of
Arce et al.~(\citeyear{arce}) for an expanding bubble (see their
Fig.~5 and Sect.~3).
In order to investigate this aspect in more detail, 
we computed position-velocity (p-v) plots along perpendicular
cuts marked in the left panel of Fig.~\ref{pv} centred on the infrared source.
As seen in the figure, the two position-velocity diagrams show a U-like
shape. In the Arce et al.'s model, an expanding shell would give a ring in the (p-v) plots
(see their Fig.~5), while we only see the redshifted half of it. However,
Arce et al.~(\citeyear{arce}) suggests that such a feature is consistent 
with a source driving the bubble slightly displaced 
towards the observer with respect to the surrounding molecular gas, so 
that we mainly see the gas which is moving away from us.
%The shell centre, when projected in a perpendicuar direction to each cut, point 
%towards the infrared source located at the centre of the green 
%circle in Fig.~\ref{pv}. 
This suggests that the infrared source is sweeping out the surrounding 
material through its winds/radiation, with an expansion velocity
(difference between the 'tip' of the U-like feature and the 'tails')
of about 1 \kms\ (see Fig.~\ref{pv}). 
In addition, the largest \AMM\ (1,1) line width in the region is found very 
close to this infrared source (see Figure~\ref{moments}, right panel), suggesting 
again the interaction of its winds/radiation with the dense surrounding 
gas. Thus, the more evolved members of the protocluster 
seem to be kinematically perturbing the starless cores, like 
condensation N. This finding raises further questions about interactions 
in protoclusters, such as to what extent are the physical (temperature 
and density), and chemical (C-bearing species, \AMM\ /\H\ ratio, deuterium
fractionation, etc.) properties of starless cores affected by more evolved cluster 
members. Thus, the I05345 region harbours cores which are eminently suitable 
for assessing the influence of formed stars on starless 
condensations.
% and constitutes an ideal laboratory for further studies.

%A tentative velocity gradient in the north-south direction can also be 
%noticed towards core S. In this core, a velocity gradient induced by the wall of the red lobe of
%the outflow was already suggested in paper~II using the \D\ (3--2) line peak velocities.
%The present work marginally supports this finding and we suggest that
%this perturbation could be due to the interaction with the nearby C2 core, 
%but only deeper sensitivity and higher spectral and angular resolution observations
%can shed light on this issue. We point out that the plots in Fig.~\ref{velocity_maps} 
%have been made interpolating the input data points on a regular grid. 
%Because of this method, the values towards condensation N, 
%along the external edge of the emission 
%and in the southern portion of the main filament could not be very
%accurate because there are several points with faint emission there.

%\subsubsection{North-western filament}

\subsection{Temperature and column density}
\label{temp_maps}

\begin{figure}
\centerline{\includegraphics[angle=-90,width=4.6cm]{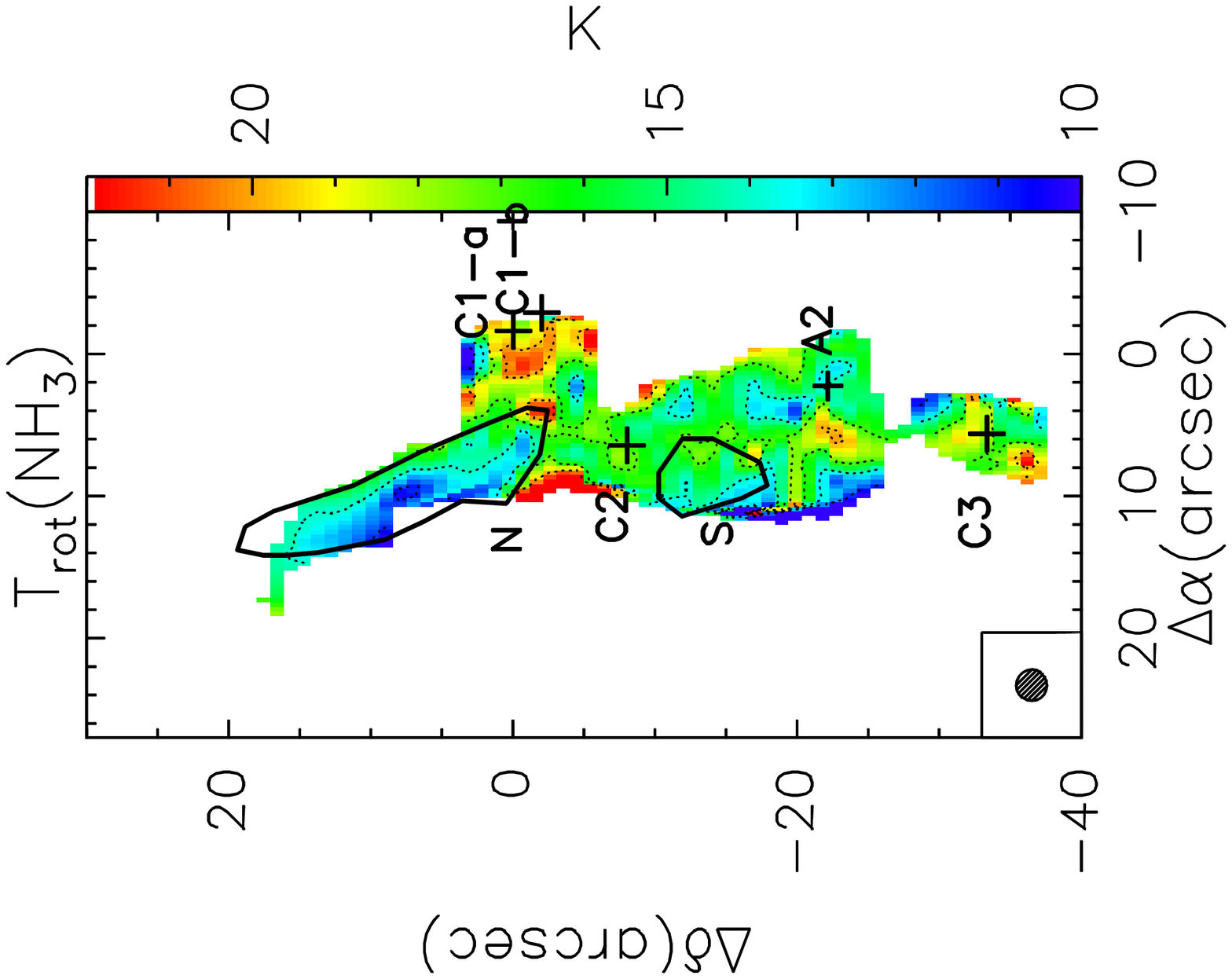}
                    \includegraphics[angle=-90,width=4.2cm]{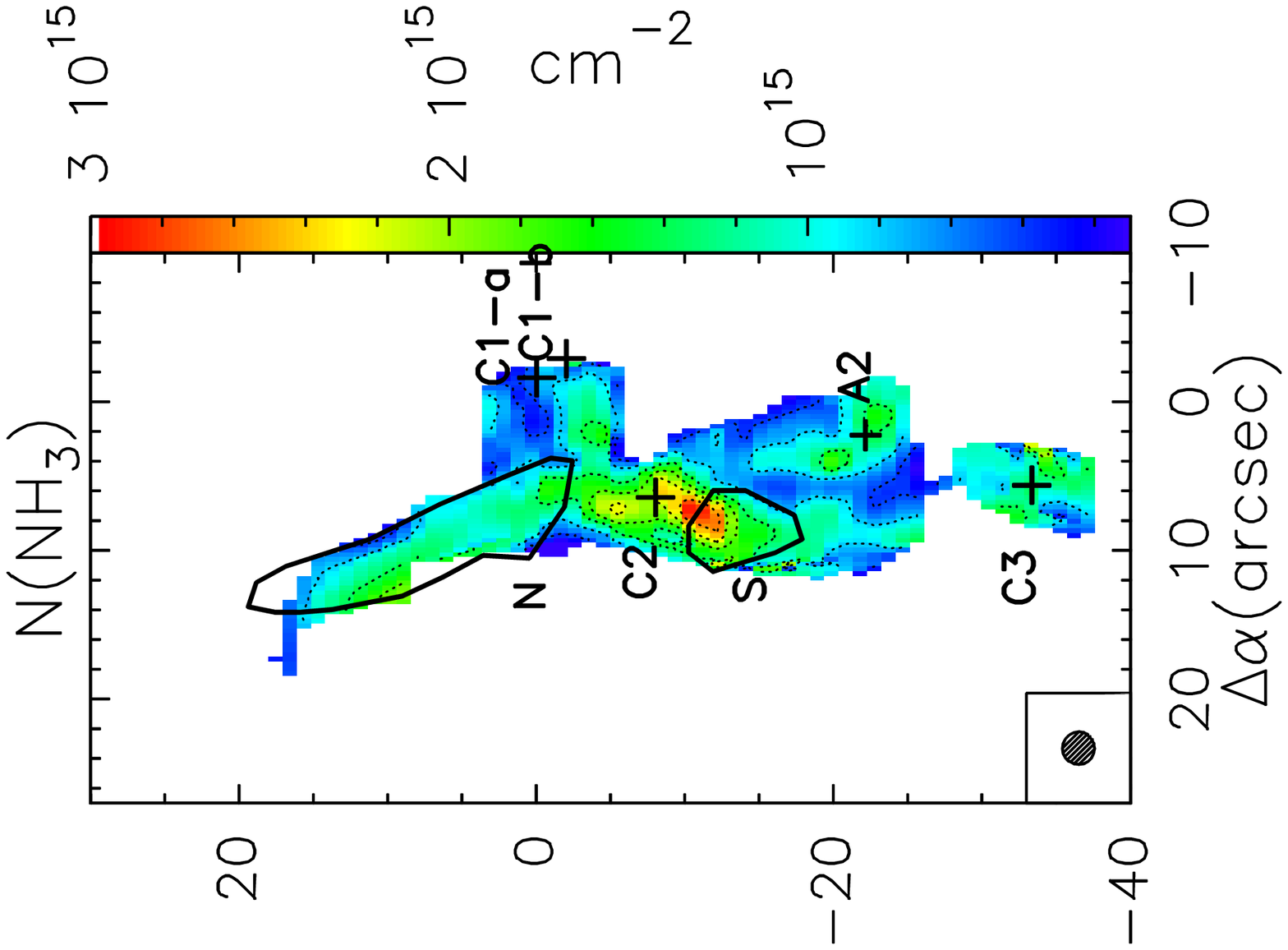}}
\caption{Map of rotation temperature (\Tr , left panel) and \AMM\ total column density 
($N$(\AMM ), right panel) derived from the \AMM\ (2,2)-to-(1,1)
ratio as described in Sect.~\ref{temperature}. In left panel, contour levels
(dotted lines) range from 11~K to 20~K in steps of 3~K,
while they range from 4$\times 10^{14}$\cmq\ to 3.2$\times 10^{15}$\cmq\
in steps of 4$\times 10^{14}$\cmq\ in the right panel.
Crosses and solid contours indicate the
millimetre continuum cores and the \D\ cores, respectively, as in Figs.~\ref{I05345_NH311_int}
and ~\ref{I05345_NH322_int}. In both panels, the ellipse in the bottom left corner shows
the VLA synthesised beam.
}
\label{temp_map}
\end{figure}

Following the procedure described in Sect.~\ref{temperature} we have
derived rotation temperature and total ammonia column density across
I05345. The results are shown in Fig.~\ref{temp_map}. 
We have excluded from the plots the arch-like filament,
in which the (2,2) transition as well as the satellites of the
(1,1) line are generally too faint to derive accurate fit results.
In general we also excluded from the analysis the pixels for which the 
(2,2) line is not detected at a $\sim 4 \sigma$ rms level, and we
have used the same interpolation method as for Fig.~\ref{velocity_maps},
thus we recommend to regard with caution especially the values
at the edge of the emission and towards N.
In the left panel we see that
\Tr\ is mostly in between 15 and 20~K along the main filament, especially close
to C1 (both C1-a and C1-b), C2 and C3, while it is lower (less than 15~K) towards 
N and in the eastern part of S. 

Although the temperature map does not reveal big gradients,
one can notice three temperature enhancements: (east of) C1-b, north-east of C2, and 
east of A2. The temperature enhancement in C1-b could be due to a 
combination of internal heating from the hot core (paper~II) 
as well as heating by the passage of the outflow, which seems to be 
launched around this position. The enhancement north-east of C2 could be 
due to either external heating by C2 (although this is not too likely 
because there is no clear heating associated with C2) or to the red lobe 
of the widespread CO outflow (see panel (b) in Fig.~\ref{amm_n2hp}). 
%External heating from the nearby mid-infrared sources detected in 
%the IRAC bands located east of the main filament can also contribute (Fig.~\ref{irac}).
However, because this temperature enhancement is found at the edge of 
the emission, where the errors on \Tr\ are of 4-5~K, we cannot be sure that this is real. 
The enhancement east of A2 could be possibly tracing an embedded source 
which would be related to the (faint) continuum emission detected at 
this position. Only high-sensitivity millimetre and infrared maps can 
help us to understand the nature of this temperature enhancement.
Note that the average temperature of the
plateau that surrounds the temperature enhancements
east of C1-b and east of A2 is 15.8~K (with a 1$\sigma$ rms of 1~K), 
i.e. significantly lower than the maxima measured (20 and 22~K,
respectively).
Therefore, the temperature enhancement proposed, 
except that north-east of C2, can be considered statistically significant.
%Finally, we 
%speculate that the apparent temperature gradient increasing from east to 
%west in the southern side of the main filament could be due to the 
%presence of either the mid-infrared sources revealed in the Spitzer
%IRAC images (see Fig.~\ref{irac}), or even to
%the stellar cluster associated to the IRAS source located further away
%(1\arcmin ) to the west of the \AMM\ emission (Fig.~\ref{2mass}).

In general, the measured temperatures across I05345 are higher 
than those derived towards infrared-dark clouds (IRDCs),
believed to be the earliest phases of massive star and cluster formation,
in which \Tr\ is of the order of $\sim 8 - 13$ K (Ragan et al.~\citeyear{ragan}).
They are also significantly higher than in clustered cores in low-mass
star-forming regions such as Perseus (Schnee et al.~\citeyear{schnee},
Foster et al.~\citeyear{foster}), where typical \Tr\ from
$\sim 9$ to $\sim 15$~K are measured. This indicates that the
presence of {\it already formed} intermediate- and high-mass
protostellar objects make the temperature of the dense gas 
systematically higher.
%We note, however, that the difference between starless and star-foming 
%cores is not dramatic ($\sim 3-5$~K on average). A similar difference is found 
%in both IRDCs (Ragan et al.~\citeyear{ragan}) and the Perseus 
%cores, in which the average difference between starless
%and star-forming condensations is of few degrees K.

The ammonia column density is on average of the
order of $\sim 10^{15}$\cmq\ across the filament
(right panel in Fig.~\ref{temp_map}), with the highest peak in
between S and C2, and significant peaks towards C3, A2,
south of N and east of C1-b. 

\subsection{Non-thermal contribution to the velocity dispersion}
\label{nonthermal}

Assuming the gas to be in Local Thermodynamic Equilibrium (LTE)
so that \Tr\ approximates the kinetic temperature, $T_{\rm k}$,
we can compute the non-thermal contribution to the observed 
velocity dispersion from the relation:
$
\sigma_{\rm nth} = \sqrt{\sigma_{\rm obs}^2- \sigma_{\rm th}^2} 
%\simeq  \sqrt{ \left( \frac{\Delta v}{\sqrt{8{\rm ln}2}}\right)^2 - \sigma_{\rm th}^2 }
$
where $\sigma _{\rm obs}$ is the measured velocity dispersion of the core 
($= \Delta v/(8{\rm ln}2)^{1/2}$ for a Gaussian line profile with $\Delta v$ the 
measured FWHM of the line) and $\sigma_{\rm th}$ is the thermal velocity disperion. 
For \AMM , assuming a Maxwellian velocity distribution, $\sigma_{\rm th}$ 
can be approximated by $\sigma_{\rm th} \simeq 0.022 \sqrt{T_{\rm k} {\rm (K)}}\;\kms $,
with $T_{\rm k} =$\Tr .
In Fig.~\ref{nonthermal} we plot both $\sigma_{\rm nth}$
and the ratio $\sigma_{\rm nth}/ \sigma_{\rm th}$,
derived from the \AMM\ (1,1) lines. As expected, the non-thermal contribution
to the line widths is more pronounced close to the centre of the star formation
activity, especially towards C2 that likely harbours an \HII\ region, but also towards 
C1 and C3. For all these cores $\sigma_{\rm nth}$
exceeds $\sigma_{\rm th}$ of a factor larger than $\sim 3.5$. Towards N, S, 
A2 and the diffuse intracluster emission of the main filament,
the ratio is mostly in between 2 and 3, suggesting that the gas is more quiescent
here but still with a dominant non-thermal contibution.
In particular, we highlight the clear drop in the ratio $\sigma_{\rm nth}/ \sigma_{\rm th}$
at the edge of core S, which resambles the sharp transition in the gas 
turbulence observed inside and out of the low-mass dense core Barnard 5 
(Pineda et al.~\citeyear{pineda10}). 

The plot is in agreement with the one made in paper~II from the \H\ (1--0)
line widths (see left panel of their Fig.~11) which, however, was limited to the
central portion of the main filament. The increase in the $\sigma_{\rm nth}$
close to YSOs is found also in low-mass star-forming regions 
(see e.g. Pineda et al.~\citeyear{pineda}), but the difference here is more dramatic.
Andr\'e et al.~(\citeyear{andre}) and Foster et al.~(\citeyear{foster}) found
that in the low-mass clustered cores embedded in Ophiuchus and Perseus 
$\sigma_{\rm nth}$ is comparable to or smaller
than the thermal broadening, respectively, regardless of whether the 
cores are isolated, clustered, starless or star-forming. 
On the other hand, in the few examples of protoclusters that contain 
more massive objects, line widths comparable to those measured
in I05345 have been found
(e.g.~Palau et al.~\citeyear{palau2007a}, Busquet et al.~\citeyear{busquet10b}, 
Pillai et al.~\citeyear{pillai11}). 

These findings suggest that on the spatial scales resolved by us, i.e. $\sim 0.02$~pc, 
the non-thermal motions are dominant in protoclusters that contain
intermediate- and high-mass objects, and the dissipation of 
turbulence does not seem to occur at these sub-pc scales. 
High levels of turbulence on small spatial scales
can be due to the overall high pressure of the environment (McKee \& Tan~\citeyear{mcKee02}) 
and/or to large amounts of energy and momentum injected locally 
in the environment by the most massive
protocluster members. The presence of the widespread \CO\ outflow
which covers a large portion of the dense gas and interacts with most of the 
condensations (see middle panel in Fig.~\ref{nh3_n2hp})
suggests that the turbulence-injected scenario is very likely for I05345.

%In paperI, we speculated that 
%the highly turbulent starless condensations N and S might be the seeds 
%of future more massive star(s) if still accreting matter.
%On the other hand, in this work we have also found
%that the ratio between gas mass and virial mass is larger than 1, 
%indicating that the condensations are likely on the verge of gravitational
%collapse (see Sect.~\ref{masses}).
%Therefore, the evolution of these starless condensations is still unclear.

\begin{figure}
\centerline{\includegraphics[angle=-90,width=5.2cm]{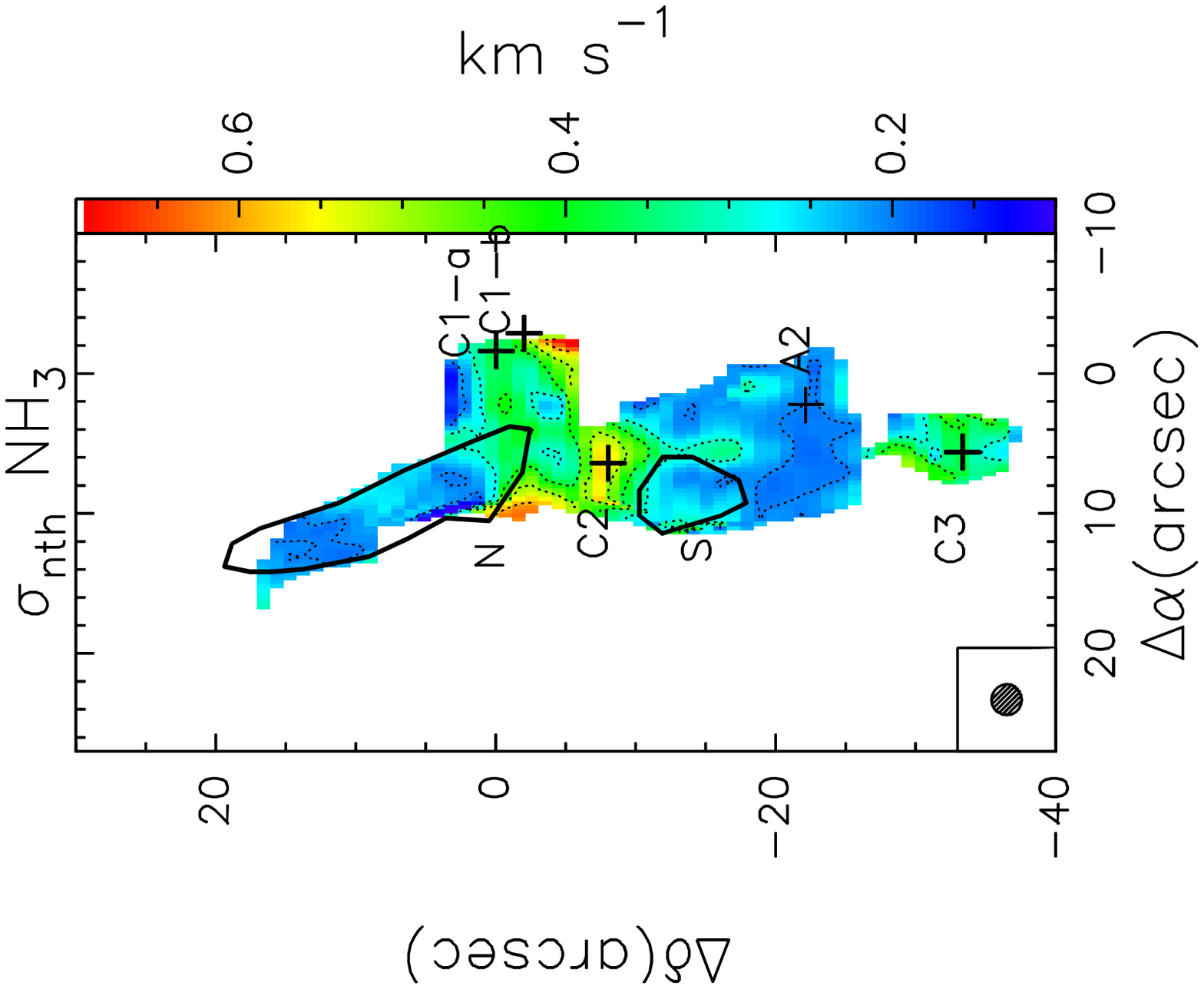}
                    \includegraphics[angle=-90,width=3.7cm]{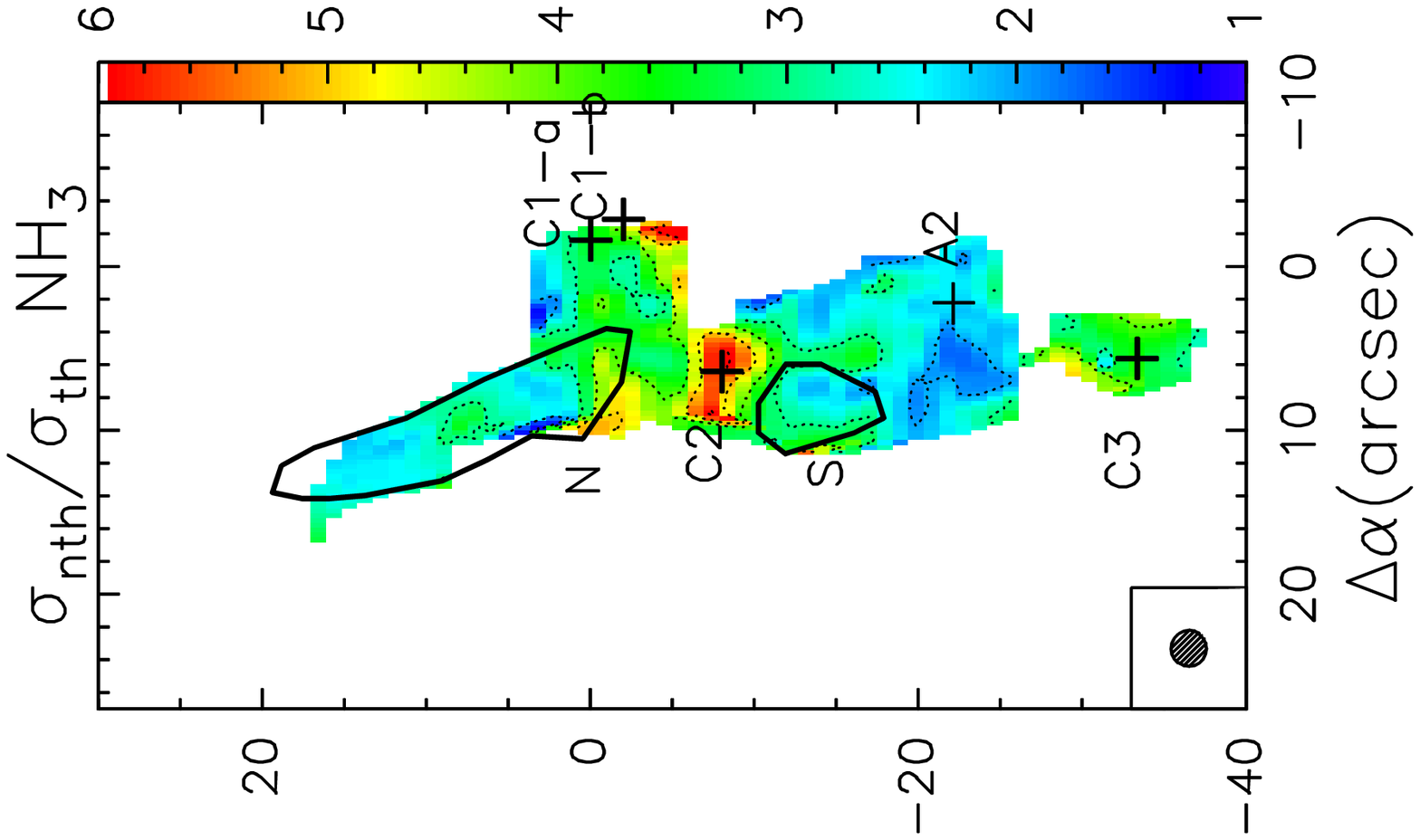}}
\caption{Map of the non-thermal velocity dispersion ($\sigma_{\rm nth}$, left panel)
and ratio between non-thermal and thermal contribution to the
observed velocity dispersion ($\sigma_{\rm nth}/\sigma_{\rm th}$, right panel). 
In the left panel, the contour levels (dotted contours) range from 0.2 to 0.6~\kms , in steps
of 0.1~\kms , while in the right panel they range from 1 to 6, in steps of 1.
Crosses and solid contours indicate the millimetre continuum cores and the \D\ cores, 
respectively, as in Figs.~\ref{I05345_NH322_int}
and ~\ref{I05345_NH311_int}.}
\label{nonthermal}
\end{figure}

\section{Discussion}
\label{discu}

\subsection{\AMM\ and \H\ relative abundance: an evolutionary tracer?}
\label{abundances}

Unlike most of the C-bearing molecules, the nitrogen-bearing
molecules \AMM\ and \H\ do not suffer from depletion 
even at high-densities (i.e. in between $\sim 10^{5}$ and 
$\sim 10^{7}$ \cmc , e.g.~Tafalla et al.~\citeyear{tafalla02}) and are therefore
excellent tracers of dense gas. Studies in both low-
and high-mass star-forming regions (Hotzel et al.~\citeyear{hotzel}, 
Palau et al.~\citeyear{palau2007a}) 
suggest that the abundance ratio of the two species in dense cores is sensitive 
to evolution during the pre-stellar phase and the very beginning of the 
protostellar phase, with the ratio \AMMtoH\ decreasing with time. Once CO is 
considerably desorbed from grain mantles by protostellar feedback 
(either by outflow passage or by an increase in temperature), \H\ is 
progressively destroyed and the ratio increases again  (Busquet et al.~\citeyear{busquet11}). 
However, the underlying physical and chemical 
reasons for the high values of \AMMtoH\ ratio during the pre-stellar 
phase and its subsequent decrease have not been investigated yet.

To probe this intriguing chemical behaviour in the cores 
of I05345, we have extracted spectra of \H\ (1--0)\footnote{from
the PdBI observations published in paper~I and using
the same integration regions as those used to extract the
ammonia spectra} and derived \H\ column densities following 
the method described in paper~I.
We excluded C3 from the analysis because this is outside the primary beam of the
\H\ observations.
In Table~\ref{nh3_n2hp} we list the resulting column density ratio 
$N$(\AMM )/$N$(\H ). 
We find ratios from $\sim 30$ to $\sim 80$ in all the protostellar objects,
$\sim 60$ and $\sim 100$ for A1 and A2, 
respectively, and $\sim 400$ and $\sim 600$ for the \D\ condensations
S and N, respectively. 
Because N and S are supposed to be pre--stellar, the relative abundance 
of \AMM\ to \H\ seems greatly enhanced in pre--stellar cores
(we point out that both A1 and A2 were close to the edge of the primary
beam of the \D , and were undetected, with a \D\ column
density upper limit of $\sim 5 \times 10^{11}$ \cmq , corresponding
to an upper limit on the \D -to-\H\ column density ratio of $\sim 0.01$). 
%This is highlighted by Fig.~\ref{masse_abbondanze}, where 
%we plot the \AMM -to-\H\ column density ratio as a function
%of $M_{\rm (NH_3)}$/\mvir\ : among the starless cores, the two 
%deuterated pre--stellar cores have the largest \AMM -to-\H\ column 
%density ratio.

%\begin{figure}
%\centerline{\includegraphics[angle=-90,width=7cm]{masse_abbondanze.eps}}
%\caption{\AMM -to-\H\ column density ratio against the ratio $M_{\rm (NH_3)}$/\mvir\
%in logarithmic scale for protostellar cores (stars), starless cores detected in \D\ (squares)
%and starless cores undetected in \D\ (triangles).}
%\label{masse_abbondanze}
%\end{figure}

That the \AMM -to-\H\ relative abundance in the gas phase is somehow enhanced 
in quiescent cores was firstly found in
the intermediate-/high-mass protocluster IRAS 20293+3952
by Palau et al.~(\citeyear{palau2007a}),
in consistency with observations of low-mass star-forming regions 
(Hotzel et al.~\citeyear{hotzel}, Friesen et al.~\citeyear{friesen10}).
%The interpretation of this, however, is non trivial because 
%\AMM\ is typically very abundant in warm gas, contrary to what we find in this work.  
%This is due to the fact that \AMM\ can be produced also on dust grains, 
%unlike \H\ that can form in the gas phase only, and one would expect evaporation of 
%\AMM\ from the grain mantles and a significant enhancement of its fractional 
%abundance in hot and shocked gas
%(see e.g.~Pauls et al.~\citeyear{pauls}, Bachiller~\citeyear{bachiller}).
%Also, up to now no relation between deuterated species 
%and \AMM -to-\H\ ratio was found.
In the low-mass regime, the \AMM /\H\ abundance ratio was also found to
increase towards the centre of low-mass starless cores (e.g. Tafalla et al.~2002),
which can be understood when freeze-out of species heavier than He becomes 
important (e.g. Flower et al.~\citeyear{flower}; Aikawa et al.~\citeyear{aikawa05}).
Depletion of heavy species like CO and other neutrals is also at the origin of the 
deuteration mechanism (see e.g.~Millar et al.~\citeyear{millar}) which could
then explain the relation between high deuteration and high \AMMtoH\ abundance
ratio. 
Given that no detailed explanation of the chemical processes driving the \AMM /\H\ 
increase can be found in the literature, we summarise the main passages
in Appendix-A, where we also present the predictions of a simple chemical 
model able to explain the observed \AMMtoH\ enhancement in dense cores 
characterised by high values of molecular depletion.

\begin{table}
\begin{center}
\caption[]{\AMM\ -to-\H\ column density ratio for the cores. 
In parentheses we give the error on the column densities. The uncertainty
on the ratio $N({\rm NH_3} )/N({\rm N_2H^+} )$ is of the order of the
60$\%$.}
\label{nh3_n2hp}
\begin{tabular}{ccccc}
\hline \hline
% \multicolumn{5}{c}{ \AMM\ (1,1)  \tablefootmark{a}  } \\
core &  & $N$(\H ) & $N$(\AMM )  & $\frac{N({\rm NH_3} )}{N({\rm N_2H^+} )}$ \\
  & & ($\times 10^{13}$\cmq )     & ($\times 10^{14}$\cmq ) &  \\
\hline
%sou     N(N2Hp)  sN(N2Hp)       N(NH3)               nh3/n2hp
C1 &  protostar &   1.3(0.4)  &  4(0.5) & 30 \\
C1-a &   protostar &   0.8(0.2)   & 6.1(0.7)  & 76 \\
C1-b   &   protostar & 1.2(0.3)  &   4.6(0.1)&  38 \\
C2    & protostar (\HII ?) &   2.8(0.4)   & 13(2)  & 47 \\
%C3     &  1.6(0.2)  &  27.2  & 165.6 \\
N     &  starless (\D ) & 0.10(0.06)  &  6(1)  & 595 \\
S      & starless (\D ) &  0.23(0.05) &    10(2) &   427 \\
A1  & starless (no \D )& 0.33(0.02) & 2(0.3) & 61 \\
A2  &  starless (no \D) & 0.58(0.02)  &  6(1) &  104 \\
\hline
\end{tabular}
\end{center}
\end{table}

\subsection{Implications for the formation of (proto-)clusters}
\label{nature}

An aspect with important implications for the formation of 
stellar clusters is how the observed separation between the members 
of a protocluster compares to typical fragmentation scale(s).
A typical scale is the thermal Jeans length, which can be
evaluated from the temperature according to the relation (Stahler \& Palla~\citeyear{stahler}):
\begin{equation}
\lambda_{\rm J}=\sqrt{ \frac{\pi c^{2}_{\rm s}}{G \rho} } = 0.19 {\rm pc} \left(\frac{T}{10 {\rm K}}\right)^{1/2} \left( \frac{n}{10^{4} {\rm cm^{-3}}} \right)^{-1/2}\;\;,
\end{equation}
where $c^{2}_{\rm s}$ is the sound speed, and $\rho$ and $n$ are the mass density
and volume number density, respectively. 
To compute a reasonable estimate of the Jeans length in \i\ we
consider an interval of 15-20~K for the temperature. 
For the density, the average density of the clump in which
I05345 is embedded is $\sim 1\times 10^{6}$ \cmc\ 
(Fontani et al.~\citeyear{fontani06}). With these values,
we obtain a thermal Jeans length $\lambda_{\rm J} \sim 0.023$ -- 
0.027~pc, corresponding to angular scales of $\sim 2 \farcs 5$ .
The spatial resolution of our observations is barely
sufficient to resolve these scales.

The average projected separation among the intensity peaks
of the protocluster members within the central part of the main filament 
is 10\arcsec\ (see e.g. left panel in Fig.~\ref{I05345_NH311_int}), 
larger than the expected thermal Jeans length.
The two sources separated more than the Jeans length from any other cluster 
member are A2 and C3. However, it is still unclear which is the
nature of these condensations and whether they are real protocluster members
(especially C3 which is quite isolated from the rest). If we consider only the cores that 
almost certainly belong to the protocluster, namely C1-a, C1-b, C2, N and S,
their projected separations are roughly consistent with thermal fragmentation.
On the other hand, the masses are of one order of magnitude higher than
the expected Jeans mass {(see Table~\ref{temp})}, $M_{\rm J}=\rho \lambda_{\rm J}^3$,
estimated to be between $\sim 0.1$ and $0.18$ \solm . 

Zhang et al.~(\citeyear{zhang09}) found a similar result in the 
IRDC G28.34+0.06, with core
separations comparable to the Jeans length and core masses
larger than the Jeans masses, and concluded that non-thermal
motions such as turbulence and/or magnetic fields 
can be dominant in the fragmentation of the IRDC. 
Similar findings have been found by Pillai et al.~(\citeyear{pillai11}),
who derived that the fragmentation in two IRDCs cannot be
explained by thermal fragmentation alone, but must include significant
contributions from non-thermal motions.
Because of the high non-thermal velocity dispersion measured
(Sect.~4.3), a
turbulent-driven fragmentation scenario or a significant contribution
of magnetic fields can be applied to I05345 as well.
%If we include the non-thermal contribution, which is 
%about 0.4--0.5 \kms\ (see Sect.~4.3), we obtain Jeans masses 
%around 15 \solm , which are comparable to the masses 
%measured in Table 2.
%This suggests that, if the non-thermal broadening is due to turbulence, 
%turbulence is sufficient to explain fragmentation 
%in this cloud.

\section{Summary and conclusions}
\label{summary}

We carried out VLA observations of the ammonia (1,1) and
(2,2) inversion transitions towards the protocluster I05345.
Ammonia is the best thermometer in dense gas and allows one
to derive both temperature, column density and kinematics 
of the dense gas associated with the protocluster.
The most direct results of this work are the following: 
\begin{itemize}
\item the ammonia emission is distributed in one main central filament
that encompasses all the previously detected high density cores,
and a secondary one to the NW of the protocluster centre;
\item the starless condensations have rotation
temperatures of about 15~K, while the protostellar cores are hotter
(\Tr\ in between $\sim 17$ and 22~K). The temperature across I05345 is on average 
higher (roughly a factor $\sim 1.3-1.5$) than 
that measured in both low-mass protoclusters and IRDCs using the
same diagnostics. We suggest that this is due to the feedback
of the already formed intermediate- to high-mass protostellar
objects both inside and outside the protocluster;
\item the non-thermal contribution to the observed
line widths is significantly higher than the 
thermal broadening both in the starless ($\sigma_{\rm nth}/\sigma_{\rm th}\geq 2$)
and in the protostellar ($\sigma_{\rm nth}/\sigma_{\rm th}\geq 4$)
condensations. Specifically, the starless condensations are more
turbulent than those observed in low-mass protoclusters (Ophiuchus
and Perseus). Like the temperature enhancement, turbulence can 
be increased by the nearby massive protostars and/or surrounding
more evolved stellar clusters;
\item the \AMM -to-\H\ column density ratio seems greatly
enhanced in the pre--stellar core candidates N and S, which
show \AMM -to-\H\ column density ratio a factor 10 larger than
the others. Predictions of a chemical model show that this
can be due to large freeze-out of molecular species heavier
than He in dense cores.
\item By comparing the observed core masses and separations to
the expected Jeans mass and Jeans length, we speculate that the
turbulence played an important role in the initial fragmentation of the
parent clump, confirming similar evidence found in IRDCs.
\end{itemize}

{\it Acknowledgments.}  FF is  grateful to R. Cesaroni 
for a careful reading of the manuscript and for providing useful 
comments and suggestions. AP is supported by
the Spanish MICINN grant AYA2008-06189-C03 
(co-funded with FEDER funds) and by a JAE-Doc CSIC fellowship
co-funded with the European Social Fund.
GB is funded by an Italian Space Agency (ASI) fellowship under contract 
number I/005/07/01.
Many thanks to the anonymous referee for his/her useful comments
and suggestions.

{}

%%%%%%%%%%%%%%%%%%%%%%%%%%%%%%%%%%%%%%%%
\begin{appendix}

%\begin{document}
\renewcommand{\thefigure}{A-\arabic{figure}}
\renewcommand{\theequation}{A-\arabic{equation}}
% redefine the command that creates the equation no.
\setcounter{figure}{0}  % reset counter 
\setcounter{equation}{0}  % reset counter 

\section*{Appendix A: \AMM /\H\ abundance ratio} 

The \AMMtoH\ column density ratio has been found to significantly increase 
in starless cores, in particular in those with the highest \D\ to \H\
abundance ratio (Sect.~\ref{abundances}).
This trend is reminiscent of the \AMMtoH\ abundance increase towards the
centre of low-mass starless
cores (e.g. Tafalla et al. 2002), and can be understood in terms of
freeze-out of species heavier than He
(Flower et al.~\citeyear{flower}; Aikawa et al.~\citeyear{aikawa05}).
We present here the predictions of a chemical model, given that a detailed explanation 
of the chemical processes leading to the \AMMtoH\
increase is lacking in the literature.

The freeze-out time scale ($t_{\rm freeze}$) is inversely proportional to
the density and becomes
shorter than any dynamical time scale if the volume density ($n_{\rm H}$)
is larger than
$\sim 10^4$ \cmc\ ($t_{\rm freeze} \simeq 10^9/n_H$ yr; e.g. Bergin \&
Tafalla~2007). Therefore,
in regions with $n_{\rm H}$ larger than a few times $10^4$ \cmc , atomic
and molecular species
heavier than He becomes increasingly more depleted from the gas phase. In
dark regions,
this process is only counterbalanced by nonthermal desorption due to
cosmic-rays impulsive
heating of dust grains (e.g. Hasegawa \& Herbst~\citeyear{heh}) and to
photodesorption caused
by UV photons produced by the interaction of cosmic rays with H$_2$
molecules (Shen et al.~\citeyear{shen};
Keto \& Caselli~\citeyear{kec}).
Non-thermal photodesorption however does not depend on volume density, so
freeze-out will eventually win in the densest regions ($n_{\rm H} > 10^6$\,cm$^{-3}$) of starless cores.

Volatile elements such as H, D, and He are not affected by freeze-out.
Actually,  abundances of  H$_3^+$ (and its deuterated forms), He$^+$ and
H$^+$ increase
towards dense regions, where neutral species such as O and CO (their main
destruction partners)
are mostly frozen onto dust grains (e.g. Walmsley et
al.~\citeyear{walmsley}).
How does this affect the abundance of \H\ and \AMM ?
To answer this question, one needs to consider the main formation and
destruction
routes of these species.

\begin{figure}
 \begin{center}
 \resizebox{\hsize}{!}{\includegraphics[angle=0]{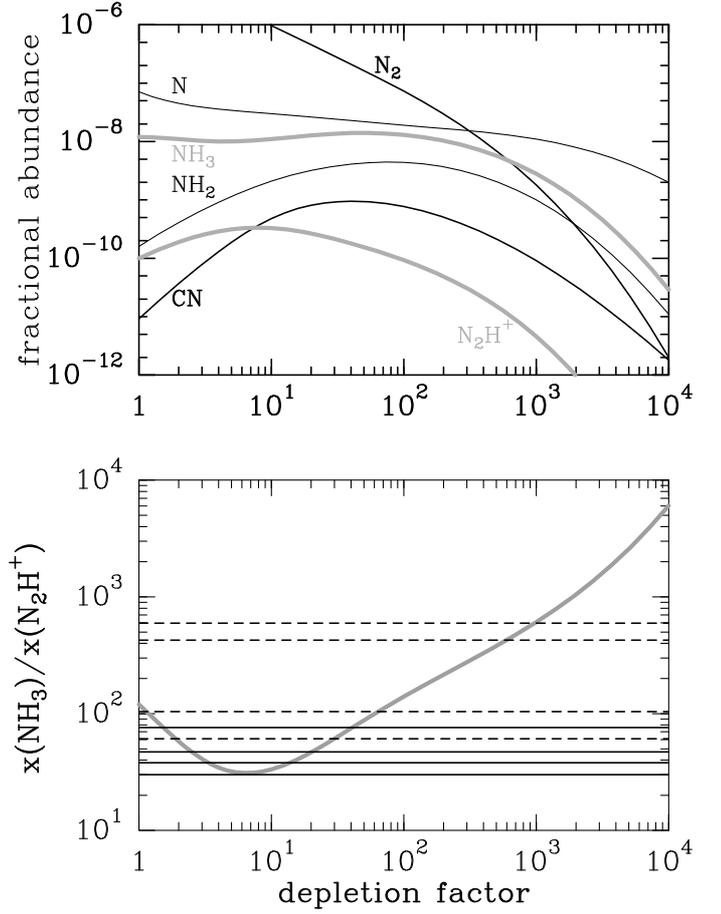}}
 \caption[]
 {\label{model}{Top panel: Fractional abundance of selected species as a
function of depletion factor
 (see text for details). Note the sharp drop of N$_2$ and the shallower
drop of \AMM\ compared to \H .
 Bottom panel: \AMMtoH\ abundance ratio as a funciton of depletion factor.
Horizontal lines are the
 observed values (see Table 3) for protostellar (continuous) and starless
(dashed) cores. Note that although
 all protostellar cores are consistent with both low and large values of
freeze-out, two of the starless cores
 can only be reproduce by values of depletion factor around 1000, similar
to what measured in the core nucleus of
 L1544 (Caselli et al. 1999). }}
 \end{center}
\end{figure}

For this purpose, we used a reduced gas-phase chemical network extracted from
KIDA\footnote{KInetic Database for Astrochemistry;
http://kida.obs.u-bordeaux1.fr/}
and run models with different amounts of elemental freeze-out.  The model
includes H,
He, O, C, N, S, Si, Fe and molecules up to 5 atoms in size. Initial
conditions are based on
standard low-metal elemental abundances (Table~1 of
Caselli~\citeyear{caselli05}), with all the elements
in atomic form, except for hydrogen which is all in H$_2$. The adopted
H$_2$ volume density,
extinction and gas temperature are 10$^5$ \cmc , 30~mag and 15~K,
respectively.
To study the effects of freeze-out, we varied the elemental abundance of
all species heavier
than He between a factor of 1 and 10$^4$. In Figure~\ref{model} (top
panel), the fractional abundace
with respect to total H nuclei of key species and (bottom panel) the \AMMtoH\ abundance
ratio ($R_{\rm N}$)
are plotted as a function of depletion factor ($f_{\rm D}$), together with
the observed values
from Table 3.  Figure~\ref{model}
shows that $R_{\rm N}$ first decreases with $f_{\rm D}$ (when $f_{\rm D}$ is
between 1 and 7) and then constantly increases to values larger than 1000.
The reason for this trend is due to the fact that \H\ is directly linked
to N$_2$,
given that its main formation route is
\begin{equation}
{\rm N_2 + H_3^+ \rightarrow N_2H^+ + H_2}\;.
\label{one}
\end{equation}
On the other hand, the formation of \AMM\ depends on N$^+$, which can be
produced by
reactions of N-bearing species such as N$_2$, CN, and NH$_2$ with He$^+$:
\begin{equation}
{\rm N_2 + He^+ \rightarrow N^+ + N + He}\;.
\label{twoa}
\end{equation}
\begin{equation}
{\rm CN + He^+ \rightarrow C + He + N^+}\;,
\label{twob}
\end{equation}
\begin{equation}
{\rm NH_2 + He^+ \rightarrow He + H_2 + N^+} \;.
\label{twoc}
\end{equation}
Reactions (\ref{twoa}), (\ref{twob}) and (\ref{twoc}) are followed by
successive hydrogen abstraction reactions
with H$_2$ until all nitrogen valence electrons are used:
\begin{equation}
{\rm N^+ \rightarrow
NH^+ \rightarrow NH_2^+ \rightarrow NH_3^+ \rightarrow NH_4^+}\;.
\label{three}
\end{equation}
Once ammonium (NH$_4^+$) is formed, it dissociatively recombines with
electrons, to form ammonia.

Reaction (\ref{twoa}) dominates in non-depleted conditions, whereas
reactions (\ref{twob}) and  (\ref{twoc})
dominate at large values of
$f_{\rm D}$, when the formation rate of N$_2$ drops. In fact, N$_2$ forms
via two (relatively slow) neutral-neutral
reactions (e.g. Hily-Blant et al.~\citeyear{hily}):
\begin{equation}
{\rm N + CH \rightarrow CN + H\;},
\label{four}
\end{equation}
\begin{equation}
{\rm CN + N \rightarrow N_2 + C\;},
\label{five}
\end{equation}
so that an increase of $f_{\rm D}$ affects the production rate of N$_2$
more severely than the production
rate of species formed via only one neutral-neutral reactions (e.g. CN,
see reaction~\ref{four}).
Moreover, as stated above, with the increase of $f_{\rm D}$, abundances of
important destruction partners of
N$_2$ (H$_3^+$ and He$^+$) increases, further reducing the N$_2$ abundance.
This is evident in the upper panel of Figure~\ref{model}, where N$_2$
shows the steepest drop in abundance
with $f_{\rm D}$.

The \H\ abundance increases between  $f_{\rm D}$ = 1 and 7, because the
drop in N$_2$ is counterbalanced
by a drop in CO, the main destruction partner at moderate values of
$f_{\rm D}$ ($\leq 10$):
\begin{equation}
{\rm N_2H^+ + CO \rightarrow HCO^+ + N_2\;}.
\label{six}
\end{equation}
At larger values of $f_{\rm D}$, dissociative recombination with electrons
becomes the most important
destruction route:
\begin{equation}
{\rm N_2H^+ + e \rightarrow N_2 + H\;(90\%)\;\;and\;\; N + NH\;(10\%)\;}.
\label{seven}
\end{equation}
Reaction (\ref{six}) reforms N$_2$, which can produce again \H\ via
reaction (\ref{one}).
Reaction (\ref{seven}) forms N$_2$ in 90\% of the cases. The other 10\% of
dissociative recombinations
form NH, permanently removing N$_2$ from the gas-phase.

Destruction of \AMM\ happens via H$_3^+$ (at large $f_{\rm D}$) and
H$_3$O$^+$ (at low $f_{\rm D}$):
\begin{equation}
{\rm NH_3 + H_3^+ \rightarrow NH_4^+ + H_2} \;,
\label{eight}
\end{equation}
\begin{equation}
{\rm NH_3 + H_3O^+ \rightarrow NH_4^+ + H_2O}\;.
\label{nine}
\end{equation}
However, unlike \H , the destruction of NH$_3$ at all values of $f_{\rm
D}$ reforms ammonium,
 the precursor of \AMM . Therefore, the \AMM\ abundance drop follows the
abundance of N$^+$
 precursors (see reactions \ref{twoa}, \ref{twob} and \ref{twoc}).  From
Figure~\ref{model}, it is evident that
 CN,  NH$_2$ and N (important for the formation of CN, see
reaction~\ref{four}) all maintain large
 abundances until $f_{\rm D} \sim 1000$, whereas \H\ shows a constant drop
for $f_{\rm D} > 10$.

\end{appendix}


\begin{thebibliography}{}

\bibitem[Aikawa et al.(2005)]{aikawa05}
Aikawa, Y., Herbst, E., Roberts, H., \& Caselli, P.~2005, ApJ, 620, 330 
\bibitem[Andr\'e et al.(2007)]{andre}
Andr\'e, P., Belloche, A., Motte, F. \& Peretto, N.~2007, A\&A, 472, 519 
\bibitem[Arce et al.(2011)]{arce}
Arce, H.G., Borkin, M.A., Goodman, A.A., Pineda, J.E., Beaumont, C.N.~2011, ApJ, 742, 105
%\bibitem[Aikawa et al.(2003)]{aikawa}
%Aikawa, Y., Ohashi, N., \& Herbst, E. 2003, ApJ, 593, 906
\bibitem[Bachiller(1996)]{bachiller}
Bachiller, R.~1996, ARA\&A, 34, 111
\bibitem[Bergin \& Tafalla(2007)]{bergin}
Bergin, E.A. \& Tafalla, M., 2007, ARA\&A, 45, 339 
%\bibitem[Bertoldi \& McKee(1992)]{bertoldi}
%Bertoldi F. \& McKee C.F.~1992, ApJ, 395, 140
%\bibitem[Beuther \& Henning(2009)]{beuther09}
%Beuther, H. \& Henning, Th.~2009, A\&A, 503, 859
\bibitem[Beuther et al.(2005)]{beuther}
Beuther, H., Thorwirth, S., Zhang, Q., et al.~2005, ApJ, 627, 834
\bibitem[Beuther et al.(2002)]{beuther02}
Beuther, H., Schilke, P., Menten, K.M.~2002, ApJ, 566, 945
\bibitem[Busquet et al.(2011)]{busquet11}
Busquet, G., Estalella, R., Zhang, Q. et al.~2011, A\&A, 525, 141
\bibitem[Busquet(2010a)]{busquet10a}
Busquet, G. 2010a, Ph.D. Thesis, University of Barcelona
\bibitem[Busquet et al.(2010b)]{busquet10b}
Busquet, G., Palau, A., Estalella, R.~et al.~2010b, A\&A, 517, L6
\bibitem[Busquet et al.(2009)]{busquet09}
Busquet, G., Palau, A., Estalella, R. et al.~2009, A\&A, 506, 1183
\bibitem[Caselli(2005)]{caselli05}
Caselli, P. 2005, in Cores to Clusters: Star Formation with Next Generation Telescopes, ed. M. S. Nanda Kumar, M. Tafalla \& P. Caselli (New York: Springer), 47
\bibitem[Caselli et al.(1999)]{caselli99}
Caselli, P., Walmsley, C.M., Tafalla, M., Dore, L., Myers, P.C.~1999, ApJ, 523, L165
\bibitem[Flower et al.(2006)]{flower}
Flower, D.~R., Pineau Des For{\^e}ts, G., \& Walmsley, C.~M.\ 2006, A\&A, 456, 215 
%\bibitem[Caselli \& Myers(1995)]{caselli95}
%Caselli, P. \& Myers, P.C.~1995, ApJ, 446, 665
%\bibitem[Caselli et al.(2002)]{caselli02}
%Caselli, P., Walmsley C.M., Zucconi, A. et al. 2002b, ApJ, 565, 344
%\bibitem[Crapsi et al.(2005)]{crapsi}
%Crapsi, A., Caselli, P., Walmsley, C.M., et al. 2005, ApJ, 619, 379
%\bibitem{evans}
%Evans, N.J.II, Shirley, Y.L. \& Mundy, L.G. 2001, ApJ, 557, 193
%\bibitem[Fontani et al.(2004)]{fontani04}
%Fontani, F., Cesaroni, R., testi, L.~et al.~2004, A\&A, 424, 179
\bibitem[Fontani et al.(2006)]{fontani06}
Fontani, F., Caselli, P., Crapsi, A., et al. 2006, A\&A, 460, 709
\bibitem[Fontani et al.(2008)]{fontani08}
Fontani, F., Caselli, P., Bourke, T.L., Cesaroni \& R., Brand, J. 2008, A\&A, 477, L45, paper~I
\bibitem[Fontani et al.(2009)]{fontani09} 
Fontani, F., Zhang, Q., Caselli, P. \& Bourke, T.L.~2009, A\&A, 499, 233, paper~II
\bibitem[Forbrich et al.(2009)]{fobrich}
Forbrich, J., Lada, C.J., Muench, A.A., Alves, J. \& Lombardi, M.~2009, ApJ, 704, 292
\bibitem[Foster et al.(2009)]{foster}
Foster, J.B., Rosolowsky, E.W., Kauffmann, J., et al.~ 2009, ApJ, 696, 298
\bibitem[Friezen et al.(2010)]{friesen10}
Friesen, R.K., Di Francesco, J., Shimajiri, Y. \& Takakuwa, S.~2010, ApJ, 708, 1002
\bibitem[Friesen et al.(2009)]{friesen}
Friesen, R.K., Di Francesco, J., Shirley, Y.L. \& Myers, P.C.~2009, ApJ, 697, 1457
\bibitem[Harju et al.(1993)]{harju}
Harju, J., Walmsley, C.M. \& Wouterloot, J.G.A.~1993, A\&AS, 98, 51
\bibitem[Hasegawa \& Herbst(1993)]{heh}
Hasegawa, T.~I., \& Herbst, E.~1993, MNRAS, 261, 83 
\bibitem[Hily-Blant et al.(2010)]{hily}
Hily-Blant, P., Walmsley, M., Pineau Des For{\^e}ts, G., \& Flower, D.~2010, A\&A, 513, A41 
\bibitem[Ho \& Townes(1983)]{ho}
Ho, P.T.P. \& Townes, C.H.~1983, ARA\&A, 21, 239
\bibitem[Hotzel et al.(2004)]{hotzel}
Hotzel, S., Harju, J., \& Walmsley, C. M. 2004, A\&A, 415, 1065
\bibitem[Keto \& Caselli(2010)]{kec}
Keto, E., \& Caselli, P.~2010, MNRAS, 402, 1625 
\bibitem[Klein et al.(2005)]{klein}
Klein, R., Posselt, B., Schreyer, K., Forbrich, J. \& Henning, Th.~2005, ApJS, 161, 361
\bibitem[Klessen \& Mac Low(2009)]{kem}
Klessen, R.S. \& Mac Low, M.-M.~2009, in "Structure Formation in Astrophysics", edited by Gilles Chabrier, ed.
Cambridge University Press, Cambridge, UK, 2009, p.427
\bibitem[Lee et al.(2011)]{lee}
Lee, K.I., Looney, L.W., Klein, R. \& Wang, S.~2011, MNRAS, 415, 2790
\bibitem[Leurini et al.(2007)]{leurini}
Leurini, S., Beuther, H., Schilke, P., et al.~2007, A\&A,  475, 925
%\bibitem[Looney et al.(2003)]{looney}
%Looney, L.W., Mundy, L.G., Welch, W.J.~2003, ApJ, 592, 255
%\bibitem[MacLaren et al.(1988)]{maclaren}
%MacLaren, I., Richardson, K.M. \& Wolfendale, A.W.~1988, ApJ, 333, 821
\bibitem[McKee \& Tan(2002)]{mcKee02}
McKee, C. F., \& Tan, J. C. 2002, Nature, 416, 59
\bibitem[McKee \& Tan(2003)]{mcKee}
McKee, C.F. \& Tan, J.C.~2003, ApJ, 585, 850 
\bibitem[Millar et al.(1989)]{millar}
Millar, T. J., Bennett, A., \& Herbst, E. 1989, ApJ, 340, 906
\bibitem[Molinari et al.(2002)]{mol02}
Molinari, S., Testi, L., Rodriguez, L.F. \& Zhang, Q.~2002, ApJ, 570, 758
%\bibitem{myers}
%Myers, M.C.~1983, ApJ, 270, 105
\bibitem[Nakamura \& Li(2007)]{nakamura}
Nakamura, M. \& Li, Z.-Y., ApJ, 656, 721 
\bibitem[Palau et al.(2007a)]{palau2007a}
Palau, Aina, Estalella, R., Ho, P.T.P., Beuther, H. \& Beltr\'an, M.T.~2007a, 474, 911
\bibitem[Palau et al.(2007b)]{palau07b}
Palau, Aina, Estalella, R., Girart, J.M., et al.~2007a,  A\&A, 465, 219
\bibitem[Peters et al.(2011)]{peters}
Peters, T., Banerjee, R., Klessen, R.S., Mac Low, M.-M.~2011, ApJ, 729, 72
%\bibitem[Pauls et al.(1983)]{pauls}
%Pauls, A., Wilson, T.L., Bieging, J.H. \& Martin, R.N.~1983, A\&A, 124, 23
\bibitem[Pillai et al.(2011)]{pillai11}
Pillai, T., Kauffmann, J., Wyrowski, F.~et al.~2011, A\&A, 530, 118
\bibitem[Pineda et al.(2011)]{pineda}
Pineda, J.E., Goodman, A.A., Arce, H.G.~et al.~2011, arXiv:1106.5474
\bibitem[Pineda et al.(2010)]{pineda10}
Pineda, J.E., Goodman, A.A., Arce, H.G. et al.~2010, ApJ, 712, L116
\bibitem[Ragan et al.(2011)]{ragan}
Ragan, S.E., Bergin, E.A. \& Wilner D.~2011, arXiv:1105.4182
\bibitem[Rathborne et al.(2008)]{rathborne}
Rathborne, J.M., Lada, C.J., Muench, A.A., Alves, J.F. \& Lombardi, M.~2008, ApJS, 174, 396
%\bibitem[Qiu et al.(2008)]{qiu}
%Qiu, K., Zhang, Q., Megeath, T. et al.~2008, ApJ, 685, 1005
\bibitem[Shen et al.(2004)]{shen}
Shen, C.~J., Greenberg, J.~M., Schutte, W.~A., \& van Dishoeck, E.~F.~2004, A\&A, 415, 203 
\bibitem[Stahler \& Palla(2005)]{stahler}
Stahler, S. W., \& Palla, F. 2005, The Formation of Stars, ed. S. W. Stahler, F. Palla (Wiley-VCH, 2005), 865
\bibitem[Tafalla et al.(2002)]{tafalla02}
Tafalla, M., Myers, P.C., Caselli, P., Walmsley, C.M. \& Comito, C. 2002, ApJ, 569, 815
\bibitem[Schnee et al.(2009)]{schnee}
Schnee, S., Rosolowsky, E., Foster, J., Enoch, M. \& Sargent, A.~2009, ApJ, 691, 1754
%\bibitem[Ungerechts et al.(1986)]{unge}
%Ungerechts, H., Walmsley, C.M. Winnewisser, G. 1986, A\&A, 157, 207
\bibitem[Varricatt et al.(2010)]{varricatt}
Varricatt, W.P., Davis, C.J., Ramsay, S. \&Todd, S.P.~2010, MNRAS, 404, 661 
\bibitem[Walmsley et al.(2004)]{walmsley}
Walmsley, C. M., Flower, D. R., Pineau des For{\^e}ts, G.~2004, 418, 1035
\bibitem[Wang et al.(2008)]{wang}
Wang, Y., ,Zhang, Q., Pillai, T., Wyrowski, F. 1\& Wu, Y.~2008, ApJ, 672L, 33
\bibitem[Ward-Thompson et al.(2007)]{wardthompson}
Ward-Thompson, D., Andr\'e, P. \& Crutcher, R.~2007, PPV conference, 951, p.~33 
\bibitem[Wu et al.(2006)]{wu}
Wu, Y., Zhang, Q., Yu, W., et al.~2006, A\&A, 450, 607
\bibitem[Zhang et al.(2005)]{zhang}
Zhang, Q., Hunter, T.R., Brand, J. et al.~2005, ApJ, 625, 864
\bibitem[Zhang et al.(2009)]{zhang09}
Zhang, Q., Wang, Y., Pillai, T. \& Rathborne, J.~2009, ApJ, 696, 268

\end{thebibliography}
\end{document}